\documentclass[twocolumn]{aastex61}
\usepackage{graphicx}
\usepackage{txfonts}
\usepackage{natbib}
\usepackage{textcomp,gensymb}
\usepackage{bm}
\shorttitle{$\Omega$-Slow Solutions and Be Stars Disks}
\shortauthors{Araya et al.}

\begin{document}

\title{$\Omega$-Slow Solutions and Be Star Disks}

\correspondingauthor{I. Araya}
\email{ignacio.araya@uv.cl}

\author{I. Araya} 
\affiliation{Instituto de F\'{\i}sica y Astronom\'{\i}a, Facultad de Ciencias, Universidad de Valpara\'{\i}so,
Av. Gran Breta\~na 1111, Casilla 5030, Valpara\'{\i}so, Chile}

\author{C. E. Jones} 
\affiliation{Department of Physics and Astronomy, The University of Western Ontario, London, Ontario N6A 3K7, Canada}

\author{M. Cur\'e} 
\affiliation{Instituto de F\'{\i}sica y Astronom\'{\i}a, Facultad de Ciencias, Universidad de Valpara\'{\i}so,
Av. Gran Breta\~na 1111, Casilla 5030, Valpara\'{\i}so, Chile}

\author{J. Silaj} 
\affiliation{Department of Physics and Astronomy, The University of Western Ontario, London, Ontario N6A 3K7, Canada}

\author{L. Cidale}
\affiliation{Instituto de F\'{\i}sica y Astronom\'{\i}a, Facultad de Ciencias, Universidad de Valpara\'{\i}so,
Av. Gran Breta\~na 1111, Casilla 5030, Valpara\'{\i}so, Chile}
\affiliation{Departamento de Espectroscop\'{\i}a, 
Facultad de Ciencias Astron\'omicas y Geof\'{\i}sicas,
Universidad Nacional de La Plata (UNLP), Paseo del Bosque S/N, 1900 La Plata, Argentina}
\affiliation{Instituto de Astrof\'{\i}sica La Plata, CCT La Plata, CONICET-UNLP, Paseo del Bosque S/N, 1900 La Plata, Argentina}

\author{A. Granada} 
\affiliation{Department of Physics and Astronomy, The University of Western Ontario, London, Ontario N6A 3K7, Canada}
\affiliation{Instituto de Astrof\'{\i}sica La Plata, CCT La Plata, CONICET-UNLP, Paseo del Bosque S/N, 1900 La Plata, Argentina}

\author{A. Jim\'enez} 
\affiliation{Centro de Investigaci\'on y Modelamiento de Fen\'omenos Aleatorios, Facultad de Ingenier\'ia, Universidad de Valpara\'iso, Valpara\'iso, Chile}



\begin{abstract}


As the disk formation mechanism(s) in Be stars is(are) as yet unknown, we 
investigate the role of rapidly rotating radiation-driven winds in this process. We implemented 
the effects of high stellar rotation on m-CAK models accounting for: the shape of the star, the oblate finite disk correction factor, and gravity darkening. For a fast rotating star, we obtain a two-component wind model, i.e., a fast, thin wind in the polar latitudes and an $\Omega$-slow, dense wind in the equatorial regions. We use the equatorial mass densities to explore  H$\alpha$ emission profiles for the following scenarios: 1) a spherically symmetric star, 2) an oblate shaped star with constant temperature, and 3) an oblate star with gravity darkening. One result of this work is that we have developed a novel method for solving the gravity darkened, oblated m-CAK equation of motion.  Furthermore, from our modeling we find a) the oblate finite disk correction factor, for the scenario considering the gravity darkening, can vary by at least a factor of two between the equatorial and polar directions, influencing the velocity profile and mass-loss rate accordingly, b) the H$\alpha$ profiles predicted by our model are in agreement with those predicted by a standard power-law model for following values of the line-force parameters: $1.5 \lesssim k \lesssim 3$, $ \, \alpha \sim 0.6$ and $\, \delta \gtrsim 0.1$, and c) the contribution of the fast wind component to the H$\alpha$ emission line profile is negligible; therefore, the line profiles arise mainly from the equatorial disks of Be stars.

\end{abstract}




\section{INTRODUCTION}
 \label{introd}
 Classical Be stars have been historically defined as ``a non-supergiant'' B star whose spectrum has, or had at some time, one or more Balmer lines in emission \citep{collins1987}. Currently, there is consensus that Be stars are very rapidly rotating and non-radially pulsating B stars (of luminosity class V-III), forming a decretion disk of outwardly flowing gas rotating in a Keplerian fashion \citep{rivinius2013}. It is generally accepted that Be star disks are geometrically thin, produced from material ejected from the central star.
        
An observational feature of these stars is the presence of emission lines from the optical to the near IR wavelength regions of the spectrum. These lines can be formed in a region very close to the star (helium and double ionized metal lines), in a large part of the disk (hydrogen lines), or relatively far from the star  \citep[singly ionized metal lines,][]{rivinius2013}. One of the predominant emission lines is the H${\alpha}$ line, and because it forms in a large region of the disk, it is often modeled to obtain average disk properties. Currently, Be star disks are frequently modeled by assuming a simple power-law density distribution in the radial direction ($\propto\,r^{-n}$) following the works of \citet{waters1986, cote1987, waters1987, jones2008, silaj2014a}. Typically, the values for the radial power-law index $n$ are found to be in the range of 2.0 to 3.5, based on fits to H$\alpha$ or the IR continuum \citep{silaj2014a,waters1986}.

Another observational feature of Be stars is the structure of their stellar winds. Observations of broad short-ward shifted UV lines of superionized elements (C\, {\sc IV}, Si\, {\sc IV} and N\, {\sc V}) demonstrate the presence of high-velocity stellar winds, even for later Be spectral types, whereas in normal B stars they are typically observed only in the early spectral types \citep{prinja1989}. Furthermore, these winds also suggest a trend of increasing $v_\infty/v_{\rm{esc}}$ (ratio between terminal velocity and escape velocity) as a function of $v\,sin\,i/v_{\rm{crit}}$ (ratio between projected rotational velocity and critical rotation speed), which is in contradiction with the radiation-driven wind theory or m-CAK theory \citep{friend1986, ppk1986}, based on the pioneering CAK theory of \cite{castor1975}, who predict a decrease in the terminal velocity as a function of rotational velocity \citep{prinja1989}. Interferometric observations have revealed some circumstellar envelopes with large scale asymmetries along the polar directions that suggest an enhanced polar wind \citep{kervella2006, meilland2007}. Polar winds from Be and normal B stars can properly be described by the current radiation-driven wind theory for massive stars \citep{rivinius2013}. In contrast to this, at the equator, Be stars present a much slower and denser mass flux which is not in agreement with the standard radiation-driven wind theory.

The precise mechanism to explain the formation of Be star disks is still under debate, and many attempts to link the radiation-driven wind theory to these disks have been made.  One such example is the Wind Compressed Disk (WCD) model proposed by \cite{bjorkman1993}. This model suggested that the wind from both hemispheres of a rapidly rotating star is deflected towards the equatorial plane producing an out-flowing equatorial disk. However, \cite{owocki1996} investigated the effects of non-radial line-forces on the formation of a WCD and their results showed that these forces (enhanced by the gravity darkening effect) may lead to an inhibition of the out-flowing equatorial disk. Later, \cite{krticka2011} examined the mechanisms of mass and angular momentum loss via an equatorial decretion disk associated with near-critical rotation. The authors emphasize the role of viscous coupling in outward angular momentum transport in the decretion disk and radiative ablation of the inner disk from the bright central star. In this context, it has been shown that radiation-driven disk-ablation models may lead to the destruction of an optically thin Be star disk on a dynamical time-scale of the order of months to years \citep{kee2016}, without invoking anomalously strong viscous diffusion. 

 Radiation-driven ablation viscous disk models are currently based on a non-rotating m-CAK wind solution that assumes a power-law distribution in line strength.
 It is then important to stress that when the finite disk approximation for the line acceleration \citep[m-CAK theory,][]{friend1986, ppk1986} is combined with the term of the centrifugal force on a rapidly rotating star  (i.e. for $\Omega \geq 75\%$ of the critical rate, with $\Omega=v/v_{\rm{crit}}$), they cause the termination of the m-CAK wind solution (fast solution) in the equatorial plane and a new solution, the $\Omega$-slow solution, is established \citep{cure2004}.

 The $\Omega$-slow solution results in higher mass-loss rates and lower terminal velocities than those from a conventional fast wind solution, so it could provide proper physical conditions to form a dense disk in Keplerian rotation via angular momentum transfer.
 
 Our model is then based on the assumption that as main sequence B stars evolve, with moderately rapid initial rotation and mass loss, they can bring angular momentum to the surface and spin up even to critical rotation \citep{ekstrom2008}. Under this condition, stars rotating near the critical rotation speed may develop a latitude-dependent wind density structure and a dense decretion disk via transfer of angular momentum. We will assume that the governing processes of a Be  star wind at high latitudes are the same as in the m-CAK  wind (fast solution). 

 At  equatorial  latitudes, the rotation term in the equation of motion will increase the mass-loss rate and decrease the terminal velocity. However, as the centrifugal term ($\propto v^{2}_{\rm{rot}}(\theta)/r^3$) increases with latitude, because $v_{\rm{rot}}(\theta)$ has larger values than $\sim 75\%$ of the critical speed, the fast solution no longer exists and the $\Omega$-slow wind solution arises. This abrupt change in the wind regime naturally produces a two-component wind. 

It is worthwhile investigating if the resulting density structure injected via the $\Omega$-slow solution for a fast rotating star is able to reproduce the H$\alpha$ emission line observed in Be stars. 

To test this hypothesis, \cite{silaj2014b} constructed a set of models for a spherical star using {\it{only}} density distributions coming from the $\Omega$-slow solution and computed the  H$\alpha$ line using the code \textsc{Bedisk} \citep{sigut2007}. Then they compared the resulting H$\alpha$ line with synthetic line profiles computed with the ad-hoc power-law model described above \citep[see, e.g.,][]{silaj2014a}. In these models, line-force parameters were taken as free parameters. These authors found that the density distribution produced by the $\Omega$-slow solution can explain the structure of a Be star disk when (unphysically) high values for the line-force parameter $k$ are assumed ($k$= 4.0, 5.0, and even $\sim$ 9.0), in contrast to the typical $k$ values self-consistent calculated ($k \lesssim 0.5$) for the fast solutions \citep[see e.g.,][]{abbott1982,ppk1986}.
In this work, we extend the study done by \cite{silaj2014b} using broader combinations of line-force parameters, and we discuss the effects of the star's oblate geometry and gravity darkening. 
Therefore, we will consider the following scenarios: a) a spherically symmetric central star with constant temperature; b) an oblate central star with constant temperature and c) an oblate central star with gravity darkening. In this framework, we also show that the contribution to the line emission coming from other latitudes above and below the equator, where the m-CAK solution governs the outflowing wind, can be neglected.

The paper is organized as follows: Section \S 2 presents our model approximations and describes the theory for the radiation-driven stellar winds, deriving the equations that account for the oblateness of the star, the oblate finite disk correction factor and gravity darkening effect. In addition, we present an overview of an ad-hoc power-law model usually used to describe the disk density structure of Be stars. In Section \S 3, we build, using the hydrodynamic code \textsc{Hydwind}, a grid of models with different values of the line-force parameters and the equatorial density structures calculated for this grid are used as input in the \textsc{Bedisk} code to obtain a grid of H$\alpha$ line profiles. This H$\alpha$ line grid is compared with the synthetic line profiles computed from the ad-hoc disk density scenario that follows a power-law distribution. Section \S 4 summarizes the results of the hydrodynamic models and the emergent line profiles predicted from the grid of equatorial mass density. In addition, we show that the contribution of the fast wind component to the emergent emission H$\alpha$ line profile is negligible. Finally, we discuss future perspectives.

%
\section{Hydrodynamical Model}

\subsection{Model approximations}
 Throughout this work we adopt the following approximations to describe the wind from a massive star with a large rotation rate:
  
\begin{itemize}

\item In this two-component stationary and isothermal wind model, we neglect the effects of viscosity, heat conduction and  magnetic fields. At the polar regions, the wind is described by the fast wind solution from the standard m-CAK model while at the equator, due to the high rotational speed, the outflowing disk-like wind is described by the $\Omega$-slow solution.
\item The hydrodynamic wind equations are solved in the 1D m-CAK model, only for polar and equatorial directions. The wind regime for other latitudes needs to be calculated using a 2D model that takes into account all non-radial  forces \citep[see e.g.,][]{bjorkman1993,cranmer1995,petrenz2000} and this is beyond the scope of this work. 
\item Line force parameters for the fast solutions correspond to the calculated values \citep[see e.g.,][and references therein]{lamers1999}. As the line-force parameters for the $\Omega$-slow solution have so far no self-consistent calculations, we adopt ranges: $0.5 \leq \alpha \leq 0.7$ and $0 \leq \delta \leq 0.2$ \citep[see][]{kudritzki2002}, while $k$ is varied within  a wider range.
\item The m-CAK model with rotation assumes angular momentum conservation. 
\end{itemize}

In the following sections, we incorporate the effect of the distortion of the star's shape caused by its high rotational speed and the gravity darkening effect into the radiation-driven wind theory.
 
\subsection{Radiation Driven Wind Equations}

The 1D m-CAK hydrodynamical equations for rotating radiation-driven winds, namely mass conservation and radial momentum conservation, considering spherical symmetry, and neglecting the gravity darkening effect and non-radial velocities are:

\begin{equation}
\label{continuity}
 r^{2}\, \rho \, v = \frac{F_\mathrm{m}}{4\, \pi}, 
\end{equation}
\noindent and

\begin{equation}
\label{momentum}
v \, \frac{dv}{dr}=-\frac{1}{\rho}\frac{dp}{dr} - \frac{G\, M \, (1-\Gamma_{\mathrm{E}})}{r^{2}} + \frac{v_{\phi}^{2}(r)}{r} + g_{\mathrm{rad}}^{\mathrm{line}}(\rho, dv/dr, n_{\mathrm{E}}),
\end{equation}

\noindent where $r$ is the radial coordinate, $F_\mathrm{m}$ is the local mass-loss rate, $v$ is the fluid radial velocity, $dv/dr$ is the velocity gradient, and g$^{\mathrm{line}}$ is the radiative acceleration as a function of  $\rho$ and $n_{\mathrm{E}}$, the mass density and electron number density, respectively. 
$\Gamma_{\mathrm{E}}$, the Eddington factor, is the Thomson electron scattering force divided by the gravitational force. The gas pressure, $p$, is given in terms of the sound speed, $a$, for an ideal gas as $p=a^2 \, \rho$. The variables $v$, $p$ and $\rho$ are a function of position and co-latitude angle. In addition, the variables $a$ and $\Gamma_{\mathrm{E}}$ become functions of co-latitude when gravity darkening is taken into account.
The rotational speed $v_{\phi}=v_{\mathrm{rot}}(\theta)\,R_{*}/r$ 
is calculated assuming conservation of angular momentum, where $v_{\mathrm{rot}}(\theta)$ is the star's surface rotational speed at co-latitude $\theta$, expressed by:

\begin{equation}
v_{\mathrm{rot}}(\theta) =  v_{\mathrm{rot}}(\mathrm{eq})\, \sin \theta
\end{equation}

\noindent with $v_{\mathrm{rot}}(\mathrm{eq})$ being the stellar rotational speed at equator.

The CAK theory assumes that the radiation emerges directly from the star (as a point source) and multiple scatterings in different lines are not taken into account. A later improvement to this theory (m-CAK) considers the radiation emanating from a stellar disk, and therefore, we adopt the m-CAK standard parametrization for the line-force term, following the descriptions of  \citet{abbott1982}, \citet{friend1986}  and \citet{ppk1986}, namely:

\begin{equation}
\label{gline}
g_{\mathrm{rad}}^{\mathrm{line}} = \frac{C}{r^{2}}\, f_{\mathrm{SFD}}(r,v,dv/dr)\, \left( r^{2} \, v \, \frac{dv}{dr} \right)^{\alpha} \left( \frac{n_{\mathrm{E}}}{W(r)} \right)^{\delta},
\end{equation}

\noindent  where the coefficient (eigenvalue) $C$ is given by 
\begin{equation}
C= \Gamma_{\mathrm{E}} \, G \, M \, k \left( \frac{4\pi}{\sigma_{\mathrm{E}} \, v_{\mathrm{th}} \, F_\mathrm{m}} \right)^{\alpha},
\end{equation} 
\noindent here $F_\mathrm{m}$ is calculated once the eigenvalue is obtained. In these equations, $W(r)$ is the dilution factor, $n_{\mathrm{E}}$ is given in units of $10^{-11}\, \rm{cm^{-3}}$, $v_{\mathrm{th}}$ is the mean thermal velocity of the protons and $ \Gamma_{\mathrm{E}}$ is expressed as
\begin{equation}
\label{gammaE}
\Gamma_{\mathrm{E}} = \frac{\sigma_{\mathrm{E}}\, L}{4\, \pi \, c \, G \, M},
 \end{equation}
 \noindent where $\sigma_{\mathrm{E}}$ is the electron scattering opacity per unit mass. 
 

The line-force parameters \citep{abbott1982,puls2000} are given by $\alpha$ (the ratio between the line-force from optically thick lines and the total line-force), $k$ (which is related to the number of lines effectively contributing to the driving of the wind) and $\delta$ (which accounts for changes in the ionization throughout the wind). \\
The finite disk correction factor, $f_{\mathrm{SFD}}$, for a spherical star, is defined as:

\begin{equation}
f_{\mathrm{SFD}}=\frac{(1+\sigma)^{1+\alpha}-(1+\sigma\mu_{*}^{2})^{1+\alpha}}{(1+\alpha)\,\sigma\,(1+\sigma)^\alpha\,(1-\mu^{2}_{*})}
\end{equation} 

\noindent where

\begin{equation}
\mu_{*} \equiv \sqrt{1-\frac{R_{*}^{2}}{r^{2}}} \:\: \mathrm{and} \:\: \sigma \equiv \frac{r}{v}\frac{d v}{d r}-1.
\end{equation}

\noindent All other variables have their standard meaning. \citep[For detailed derivations and definitions of variables, constants and functions, see, e.g.,][]{cure2004}. \\

In addition, we can define the normalised stellar angular velocity as

\begin{equation}
\label{omega1}
\Omega = \frac{v_{\mathrm{rot}}(\mathrm{eq})}{v_{\mathrm{crit}}},
\end{equation}

\noindent where $v_{\mathrm{crit}}$ is the critical rotational speed for a spherical star, defined, e.g., by \citet{maeder2000a} as:

\begin{equation}
\label{vcrit1d}
v_{\mathrm{crit}}=\left(\frac{G\, M }{R_{*}}(1-\Gamma_{\kappa})\right)^{1/2},
\end{equation}
\noindent where $\Gamma_{\kappa}$ considers the total flux mean opacity, $\kappa$, instead of the electron scattering opacity per unit mass used in the theory of radiatively driven winds given in Equation \ref{gammaE}. Because of the difficulty of knowing the exact value of $\kappa$, our values for  $v_{\mathrm{crit}}$ are calculated assuming $\Gamma_{\kappa}\cong \Gamma_{\mathrm{E}}$. This approximation represents a minor underestimation of the value of $\Omega$, since in our models both $\Gamma_{\kappa}$ and $\Gamma_{\mathrm{E}}$ are $\ll 1$.
\subsection{Oblateness and Gravity Darkening Effects}

When we take high rotation into account, the star becomes oblate and its shape becomes roughly similar to a rotating ellipsoid. The von Zeipel theorem states that the radiative flux $\bm{\mathcal{F}}$ at some co-latitude in a rotating star is proportional to the local effective gravity, $\bm{g}_{\mathrm{eff}}$ \citep{vonzeipel1924}. This oblateness changes the local effective gravity, $\bm{g}_{\mathrm{eff}}=\bm{g}_{\mathrm{grav}}+\bm{g}_{\mathrm{rot}}$ (sum of the gravitational and centrifugal accelerations), and hence the local temperature at the stellar surface \citep[][ for shellular rotation see Maeder \citeyear{maeder1999}]{vonzeipel1924}. For this reason, when we consider both effects,  it is convenient to redefine some parameters as a function of co-latitude $\theta$ and rotational speed. 

Thus, the stars' local rotational speed is given by

\begin{equation}
v_{\mathrm{rot}}(\Omega, \theta)= \frac{R(\Omega, \theta)}{R(\Omega, \mathrm{eq})}\, v_{\mathrm{rot}}(\Omega, \mathrm{eq})\, \sin \theta,
\end{equation}

\noindent and the normalized stellar angular velocity is expressed as 

\begin{equation}
\label{omega2}
\Omega = \frac{\omega_{\mathrm{rot}}}{\omega_{\mathrm{crit}}} = \frac{v_{\mathrm{rot}}(\Omega,\mathrm{eq})}{v_{\mathrm{crit}}}\frac{R_{\mathrm{eq}}^{\mathrm{max}}}{R(\Omega, \mathrm{eq})},
\end{equation}

\noindent where, in a Roche model, $R_{\mathrm{eq}}^{\mathrm{max}}=3\,R_{\mathrm{pole}}/2$ is the maximum equatorial radius when a star is rotating at the critical velocity, i.e., $\Omega=1$ \citep[see, e.g.,][]{puls2008,muller2014}. $R_{\mathrm{pole}}$ is the polar radius and, as a first approximation, it is assumed to be independent of the rotational velocity\footnote{Nevertheless, evolutionary calculations show that $R_{\mathrm{pole}}$ depends slightly on $\Omega$ due to the small changes of internal structure \citep{maeder2009}.}. Then, the critical rotational velocity for a uniform radiation field, where the effect of gravity darkening is omitted, is expressed by

\begin{equation}
\label{vcrit2d}
v_{\mathrm{crit}}=\left( \frac{G\, M }{R_{\mathrm{eq}}^{\mathrm{max}}}(1-\Gamma_{\kappa}) \right)^{1/2} = \left( \frac{2\,G\, M }{3\,R_{\mathrm{pole}}}(1-\Gamma_{\kappa}) \right)^{1/2}.
\end{equation}

It is important to note that when gravity darkening is considered, $v_{\mathrm{crit}}$ requires special attention. \cite{puls2008} state:  ``After some controversial discussions arising from the suggestion of an $\Omega$ limit by \cite{langer1997, langer1998} which has been criticized by \cite{glatzel1998} (because of disregarding gravity darkening), \cite{maeder2000a} performed a detailed study on the issue". \cite{maeder2000a} state that some authors write the critical rotational velocity as in Equation (\ref{vcrit1d}), however they emphasize that this relation is only true if it is assumed that the brightness of the rotating star is uniform over its surface. This is in contradiction with the von Zeipel's theorem that predicts a decrease in the effect of radiation pressure at equator. Surprisingly for \cite{maeder2000a}, some authors use this relation simultaneously with the von Zeipel theorem.  \cite{maeder2000a} establish that the critical rotational velocity is reached when the total gravity $\bm{g}_{\mathrm{tot}}=0$, i.e., 

\begin{equation}
\bm{g}_{\mathrm{eff}}[1-\Gamma_{\Omega}(\theta)]=0,
\end{equation}

\noindent where $\Gamma_{\Omega}(\theta)$ is the local Eddington factor. Their study finds that for moderate values of $\Gamma_{\kappa}$ (which is our case) the critical speed can be calculated, independently from $\Gamma_{\kappa}$, from the condition $\bm{g}_{\mathrm{eff}}=0$, in agreement with \cite{glatzel1998}, as

\begin{equation}
\label{vcrit}
v_{\mathrm{crit}} = \left(\frac{G\, M}{R_{\mathrm{eq}}^{\mathrm{max}}}\right)^{1/2} =  \left(\frac{2\,G\,M}{3\, R_{\mathrm{pole}}}\right)^{1/2} \quad  \mathrm{for} \quad \Gamma_{\kappa} < 0.639.
\end{equation}
 
The stellar radius can be approximated as a function of co-latitude and rotational speed by

\begin{equation}
R(\Omega,\theta) = \frac{3 \, R_{\mathrm{pole}}}{\Omega \,\sin\theta} \cos\left( \frac{\pi + \arccos(\Omega\, \sin\theta)}{3} \right)
\end{equation}

\noindent \citep{cranmer1995,petrens1996,muller2014}.

The local effective gravity at a given co-latitude, directed inwards along the local surface normal, is given by the negative gradient of the effective potential
  
\begin{equation}
\Phi=-\frac{G\,M}{r} - \frac{1}{2}\,\omega_{\mathrm{rot}}^2\,r^2\,\sin^2(\theta)
\end{equation}
\noindent \citep{cranmer1995, maeder2000a}. Thus, the two-components of the local effective gravity in spherical coordinates are
 
\begin{equation}
g_{\mathrm{eff,r}}(\Omega, \theta) = -\frac{\partial\Phi}{\partial r}
\end{equation}

\noindent and

 \begin{equation}
g_{\mathrm{eff,\theta}}(\Omega, \theta) = -\frac{1}{r} \frac{\partial \Phi}{\partial \theta}.
\end{equation}

\noindent Then, the absolute value of the local effective gravity, is

\begin{equation}
g_{\mathrm{eff}}(\Omega, \theta)= \sqrt{g_{\mathrm{eff,r}}(\Omega, \theta)^{2}+g_{\mathrm{eff,\theta}}(\Omega, \theta) ^{2}},
\end{equation}
\noindent or 
\begin{eqnarray}
\nonumber 
g_{\mathrm{eff}} (\Omega, \theta) =   \frac{G\,M}{R_{\mathrm{pole}}^{2}} \frac{8}{27} & & \left[ \left( \frac{27}{8} \left(\frac{R_{\mathrm{pole}}}{R(\Omega, \theta)}\right)^{2} - \frac{R(\Omega, \theta)}{R_{\mathrm{pole}}} \Omega^{2} \sin^{2} \theta \right)^{2} \right. \\ 
& &\left. + \Omega^{4} \left( \frac{R(\Omega, \theta)}{R_{\mathrm{pole}}} \right)^{2} \sin^{2} \theta \cos^{2} \theta \right]^{1/2}.
\end{eqnarray}
\noindent On the other hand, to derive the dependence of $T_{\mathrm{eff}}$ as function of the co-latitude at the surface of the star, we follow the work of \cite{espinosa2011},  who assume that the radiative flux in the envelope of a rotating star can be expressed, following \cite{vonzeipel1924},  as:

\begin{equation}
\bm{\mathcal{F}}=-f(r,\theta) \bm{g}_{\mathrm{eff}}
\end{equation}                  
\noindent where $f(r,\theta) $ is a function of the position to be determined and satisfies the condition, 
\begin{equation}
\lim_{r\rightarrow 0}f(r,\theta) =\frac{L}{4\pi GM}= \eta
\end{equation}
\noindent and $\eta$ is a constant that scales the function $f$  and can be rewritten in a dimensionless form, as
\begin{equation}
f(r,\theta)=\eta F_{\Omega}(r,\theta).
\end{equation}
\noindent Then for the Roche model: $\nabla\cdot\bm{g}_{\mathrm{eff}} = 2 \omega_{rot}^{2}$, an expression for the local effective temperature can be derived, \citep[see][]{espinosa2011},

\begin{equation}
\label{teffELR}
T_{\mathrm{eff}}(\Omega, \theta) = \left[\frac{L}{4\, \pi \, \sigma_{B}\, G\, M} g_{\mathrm{eff}}(\Omega, \theta) F_{\Omega} (\theta)  \right]^{1/4},
\end{equation}

\noindent and further that  {\bf $F_{\Omega}(\theta) \propto \tan(\theta)^{-2}$}. This last Equation possesses singularities at the poles and the equator of the star. At these points, the local effective temperature is  respectively given by
\begin{eqnarray}
\nonumber 
T_{\mathrm{eff}}(\Omega, \mathrm{pole}) & =& \left[\frac{L}{4\, \pi \, \sigma_{B}\, G\, M} g_{\mathrm{eff}}(\Omega, \mathrm{pole})\right]^{1/4} \\
& & \times \, \exp\left( \frac{2}{3} \Omega^2 \left(  \frac{R_{\mathrm{pole}}}{R_{\mathrm{eq}}^{\mathrm{max}}} \right)^3 \right)^{1/4}
\end{eqnarray} 
\noindent and 
\begin{eqnarray}
\nonumber 
T_{\mathrm{eff}}(\Omega, \mathrm{eq}) & = &\left[\frac{L}{4\, \pi \, \sigma_{B}\, G\, M} g_{\mathrm{eff}}(\Omega, \mathrm{eq})\right]^{1/4} \\
& & \times \, \left( 1 - \Omega^{2}  \left(  \frac{R(\Omega,\mathrm{eq})}{R_{\mathrm{eq}}^{\mathrm{max}}} \right)^3 \right)^{-1/6}.
\end{eqnarray}
Since the effective gravity, and therefore the flux, varies over the surface of the rotating star, we still need to determine the local value of $\Gamma_{\Omega}$ for a barotropic case of a non-spherical star that can be defined in terms of the reduced mass, $M_{\bigstar}$, as follows:

\begin{equation}
\label{gammaO1}
\Gamma_{\Omega}= \frac{\kappa\, L}{4\, \pi \, c \, G \, M_{\bigstar}},
\end{equation}

\noindent with

\begin{equation}
\label{gammaO2}
M_{\bigstar} = M \left( 1 - \frac{\omega_{\mathrm{rot}}^{2}}{2\, \pi \, G\, \rho_{\mathrm{m}}} \right) ,
\end{equation}

\noindent where $\rho_{\mathrm{m}}$ is the internal average density \citep{maeder1999,maeder2000a}. It is important to note that the value of $\Gamma_{\Omega}$ is independent of latitude\footnote{This is true for a barotropic case or if $\kappa$ is the same over the entire stellar surface \citep{maeder1999}.}. 

In order to account for the oblateness of a gravity darkened star, 
we need to calculate the corresponding oblate finite disk correction factor $f_{\mathrm{OFD}}$ at a given co-latitude,  

\begin{eqnarray}
\nonumber
\label{Fofd}
f_{\mathrm{OFD}} & = & \frac{r^{2}}{\pi \, R^{2}(\Omega, \theta) \left( 1+ \sigma \right)^{\alpha} }  \int^{2\pi}_{0} \int^{\Theta'}_{0} \frac{T^4_{\mathrm{eff}}(\Omega, \theta'')}{T^4_{\mathrm{eff}}} \\ 
& & \times \,   \left( 1 + \sigma \cos^{2} \left( \theta \right) \right)^{\alpha} \cos \left( \theta \right) \sin \left( \theta \right) \frac{d\theta}{d \theta'} d\theta' d\phi.
\end{eqnarray}
The reader is referred to Figure 2 and section 4.3 from \citet{pelupessy2000} for the definition of the various angles\footnote{Note that equations (24), (32) and (35) from  \citet{pelupessy2000} differ in the sign of the $\alpha$ exponent compared to our derivation, because of a typo in their  equations.}. Here we neglect the continuum correction factor since it is not important for low luminosity stars  because $\Gamma_{\mathrm{E}}$ itself is small. 
\subsection{Solving the Hydrodynamic Wind Equations}

In order to solve the 1D hydrodynamic radiation-driven wind equations, \cite{cure2004} introduced the following change of variables:  $u=-R_{*}/r$, $w=v/a$ and $w'= dw/du$, with $a_{\mathrm{rot}}=v_{\mathrm{rot}}/a$, where $a$ is the isothermal sound speed.  Based on this auxiliary set of variables, we can write an approximate  Equation of Motion (EoM) for an oblate gravity darkened star valid for $\Gamma_{\mathrm{E}} \approx \Gamma_{\Omega} \ll 1$, where $\Gamma_{\Omega}$ is evaluated with $\sigma_{\mathrm{E}}$ instead of $\kappa$,

\begin{eqnarray}
\nonumber
\label{motion-eq}
F(u,w,w') & \equiv & \left( 1- \frac{1}{w^{2}} \right) w \frac{dw}{du} + A + \frac{2}{u} + a^{2}_{\mathrm{rot}}\, u \\
& &  - \, C' \, f_{\mathrm{OFD}} \, g(u)\, (w)^{-\delta} \left( w \frac{dw}{du} \right)^{\alpha}  = 0, 
\end{eqnarray}

\noindent where

\begin{equation}
g(u)= \left( \frac{u^{2}}{1-\sqrt{1-u^{2}}}  \right)^{\delta}.
\end{equation}

To solve this non-linear differential equation we adopt $u=-R_{\mathrm{pole}}/r$ and define now $A$ and $C'$  in terms  of $\Gamma_{\Omega}$,
\begin{equation}
A= \frac{G\, M \left[ 1- \Gamma_{\Omega}\right]}{a(\theta)^{2}R(\Omega,\theta)},
\end{equation}
\noindent and 
\begin{equation}
C' = C \left( \frac{F_\mathrm{m}\, D}{2\, \pi} \frac{10^{-11}}{a(\theta)R^{2}(\Omega,\theta)}  \right)^{\delta} \left( a(\theta)^{2}R(\Omega,\theta) \right)^{\alpha-1} ,
\end{equation}
\noindent with
\begin{equation}
C= \Gamma_{\Omega}\, G \, M \, k \left( \frac{4\pi}{\sigma_{\mathrm{E}} \, v_{\mathrm{th}} \, F_\mathrm{m}} \right)^{\alpha}
\end{equation} 
\noindent and 
\begin{equation}
D= \frac{1+Z_{\mathrm{He}}\,Y_{\mathrm{He}}}{1+A_{\mathrm{He}}\, Y_{\mathrm{He}}} \left(\frac{1}{m_{\mathrm{H}}}\right),
\end{equation}
\noindent where $Y_{\mathrm{He}}$ is the helium abundance relative to hydrogen, $Z_{\mathrm{He}}$ is the number of free electrons provided by helium, $A_{\mathrm{He}}$ is the atomic mass number of helium and $m_{\mathrm{H}}$ is the mass of the proton. 

A more general expression for the EoM valid for larger values of $\Gamma_{\Omega}$ can be found in Appendix \ref{eq-GD}. In Appendix \ref{gammaO} we show the calculation of $\Gamma_{\Omega}$.
	
In order to find a physical, continuous solution of $w$, that starts at the stellar surface and reaches infinity, it is necessary  to require the solution to pass through a critical (singular) point. Its location is obtained from the singularity condition,

\begin{equation}
\label{singularity}
\frac{\partial}{\partial w'} F(u,w,w') =0, 
\end{equation}

\noindent together with a lower boundary condition (at the stellar surface) by setting the surface mass density to a specific value,
\begin{equation}
\label{denssup}
\rho(R_{*},\theta) = \rho_{*}(\theta).
\end{equation}
\noindent  At the critical point a regularity condition must be imposed, namely,

\begin{equation}
\frac{d}{du} F(u,w,w') = \frac{\partial F}{\partial u} + 	\frac{\partial F}{\partial w} w' + 	\frac{\partial F}{\partial w' } w'' =0, 
\end{equation}

\noindent  where the last term $\partial F/\partial w' =0$, due to the singularity condition (Equation \ref{singularity}).
  
Depending on the value of $\Omega$, the solution of Equation \ref{motion-eq} leads to either fast or $\Omega$-slow wind solutions. Notice that for  high rotational velocities, the standard (or fast) solution ceases to exist \citep{cure2004}.  In both cases, we use the stationary hydrodynamic \textsc{Hydwind} code \citep{cure2004} modified to take into account the oblateness and gravity darkening effects.

\subsection{Calculation of the Gravity Darkened and Oblate Finite Disk Correction Factor}

To solve the m-CAK EoM accounting for the gravity darkening and the oblate distortion of the star that is caused by its rapid rotation, we implemented the method described by \cite{araya2011}, who introduced the $f_{\mathrm{OFD}}$ factor and its calculation.   
In view of the behavior of the $f_{\mathrm{OFD}}$  factor, it is possible to obtain an approximate expression via a sixth order polynomial interpolation in the inverse radial variable $u$, i.e., 

\begin{equation}
\label{poly-fit}
f_{\mathrm{OFD}}=Q(u)\,f_{\mathrm{SFD}}.
\end{equation}

With this structure, the different topological solutions of the m-CAK model, found by \cite{cure2004}, are maintained  by the $f_{\mathrm{SFD}}$ term, but modified by the incorporation of this $Q(u)$ polynomial. This approximation (Equation \ref{poly-fit}) allows the non--linear m--CAK  EoM to be solved in a straightforward way, instead of calculating the complicated integral of $f_{\mathrm{OFD}}$ (Equation \ref{Fofd}), which would be computationally expensive and numerically  unstable. 

In this work we solve the m-CAK EoM for the polar and equatorial directions (see Section \ref{results}), but in the calculation  the latitude-dependence of the oblate finite disk correction factor and the gravity darkening are taken into account.

A finite difference iterative numerical method, described in \cite{cure2004}, is used to solve this non--linear differential equation. This numerical approach needs to start from an initial trial velocity profile in order to iterate until convergence.\\
Therefore, our iterative strategy is as follows:
\begin{enumerate}
\item A initial velocity profile $\beta$-law is used for $v(u)$ and $w(u)$.
\item $f_{\mathrm{SFD}}$ and $f_{\mathrm{OFD}}$ are calculated from $w(u)$.
\item $Q(u)$ coefficients are calculated by fitting $f_{\mathrm{OFD}}/f_{\mathrm{SFD}}$ as function of $u$ .
\item  The EoM is solved with our approximate $Q(u)\,f_{\mathrm{SFD}}$, obtaining a new velocity profile $w(u)$.
\item  Steps 2 to 4 are repeated until convergence is reached.
\end{enumerate}

\begin{figure}[h]
\center
\includegraphics[width=3 in]{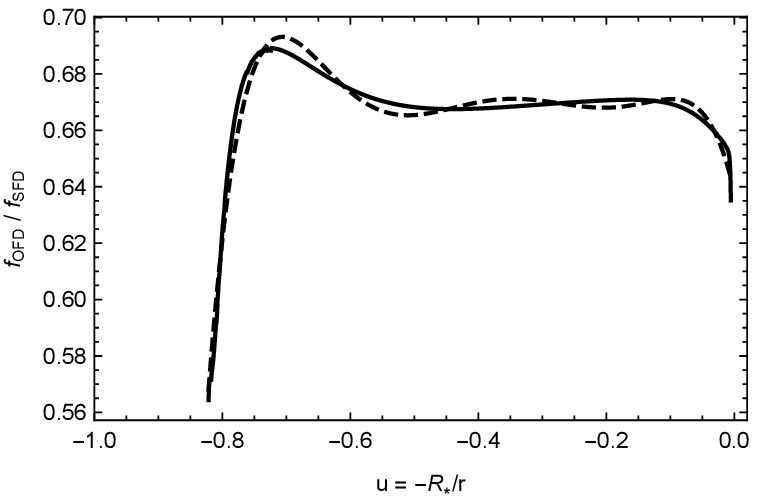}\\
\vskip 0.3cm	
\includegraphics[width=3 in]{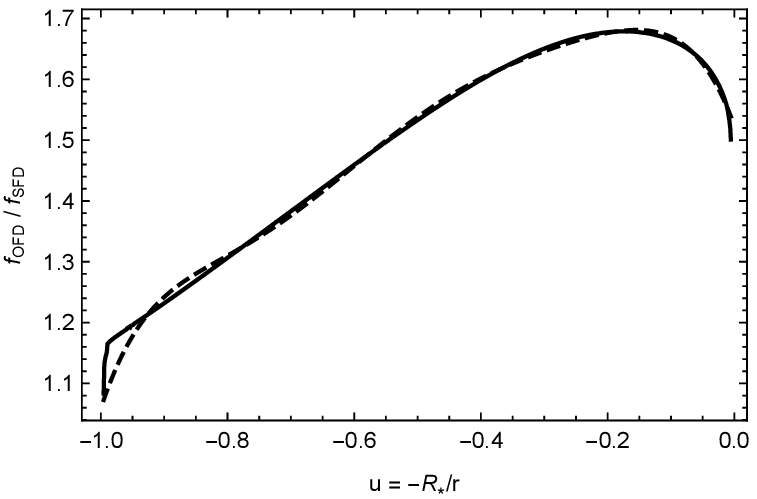}
\caption{Comparison of the ratio $f_{\mathrm{OFD}}/f_{\mathrm{SFD}}$ (solid line) with the polynomial interpolation $Q(u)$ (dashed line) at the equator  (upper panel) and the pole (lower panel). Both  expressions have been computed for a wind model solution  with $\Omega=0.90$ considering oblateness and gravity darkening effects. Note that in the upper panel the function does not start at $u=-1.0$ due to the oblateness of the star. $R_{*}$ corresponds to the polar radius of an oblate star.} 
\label{Qfit}
\end{figure}

\begin{figure}[h]
\center
\includegraphics[width=3 in]{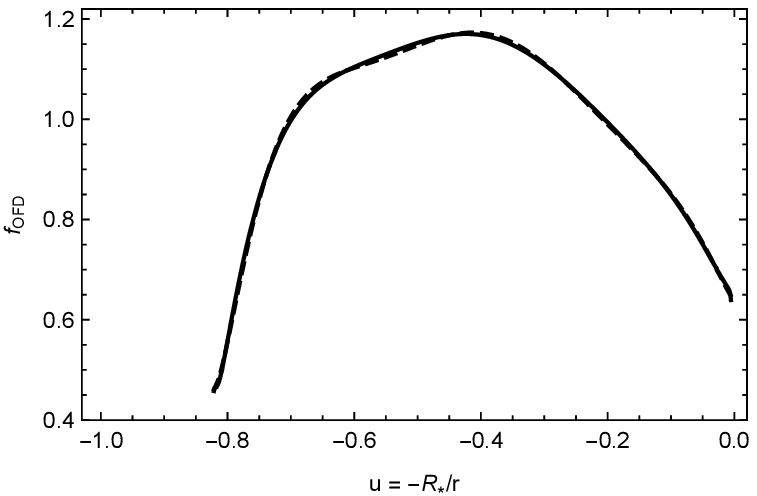}\\
\vskip 0.3cm	
\includegraphics[width=3 in]{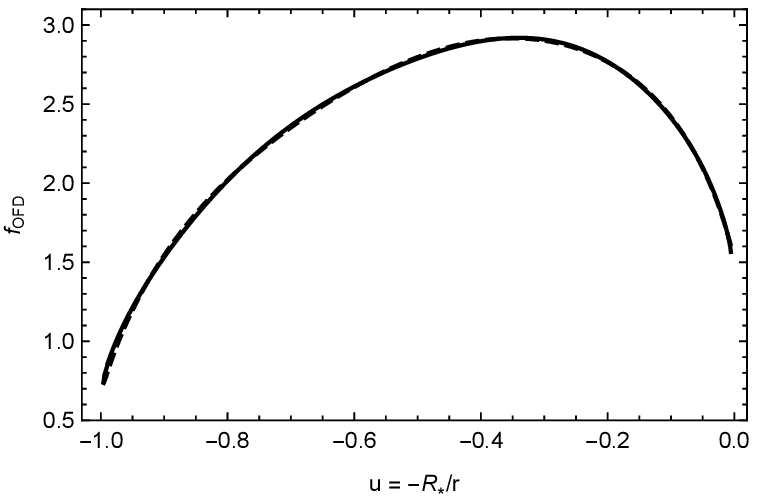}
\caption{Comparison of the oblate correction factor calculated from Equation \ref{Fofd} (solid line) with the approximate oblate correction factors obtained from our iterative method (dashed line) at the equator (upper panel) and the pole (lower panel) for a model with  $\Omega=0.90$, including gravity darkening.}
\label{ocf-plot}
\end{figure}

\begin{figure}[h]
\center
\includegraphics[width=3 in]{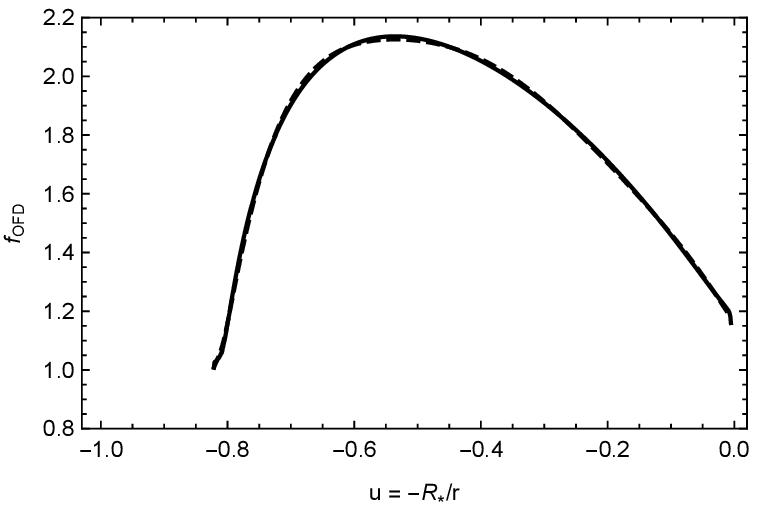}\\
\vskip 0.3cm	
\includegraphics[width=3 in]{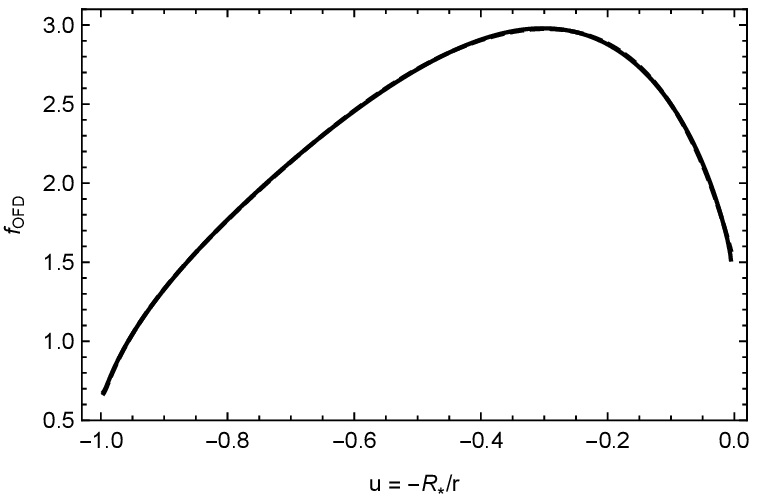}
\caption{Similar to Figure \ref{ocf-plot}, but the oblate correction factor (obtained from Equation \ref{Fofd}) is calculated at the equator (upper panel) and the pole (lower panel) without considering gravity darkening.}
\label{oblado-factor}
\end{figure}

\begin{figure}[h]
\center
\includegraphics[width=3 in]{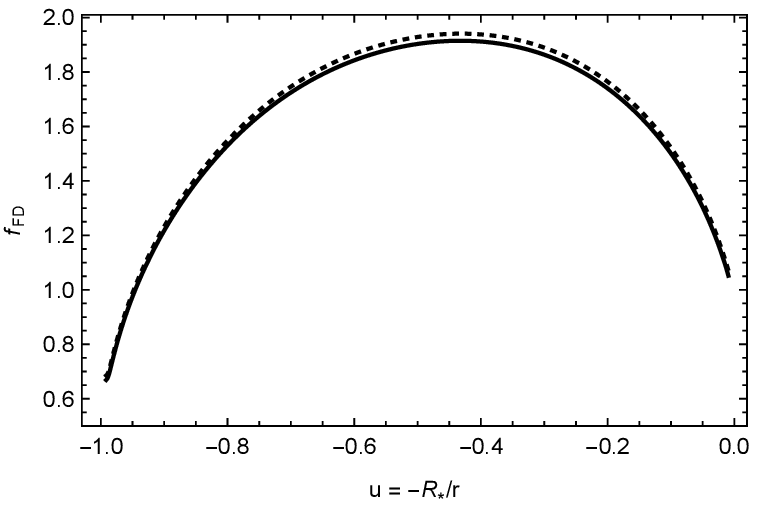}\\
\vskip 0.3cm	
\includegraphics[width=3 in]{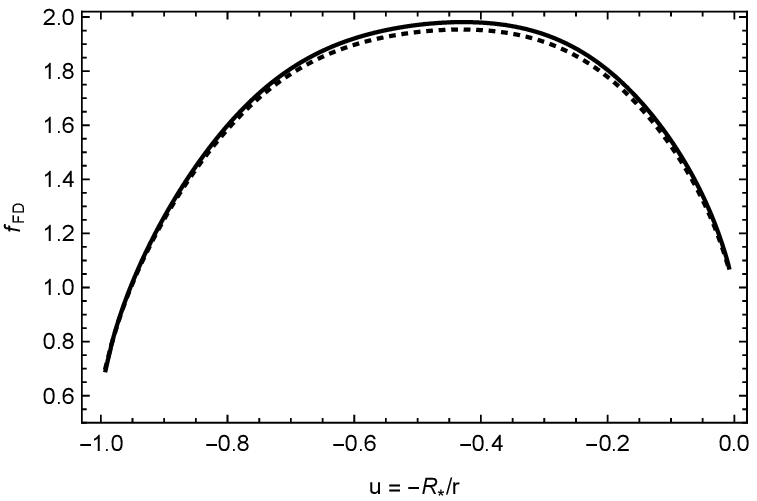}
\caption{Comparison between the $f_{\mathrm{OFD}}$ with gravity darkening (solid line) and the $f_{\mathrm{SFD}}$ (dotted line) calculated at the equator (upper panel) and the pole (lower panel) for a rotating star at $\Omega=0.2$. In both panels the correction factors are very similar.}
\label{ofd02}
\end{figure}

A comparison between $Q(u)$ and the ratio $f_{\mathrm{OFD}}/f_{\mathrm{SFD}}$ is shown in Figure \ref{Qfit} for the equatorial (upper panel) and polar (lower panel) regions. Both functions were calculated for $\Omega\,=0\,.90$ considering the oblate finite disk correction factor and gravity-darkening. We find discrepancies of about  2\%  after 4 iterations. In Figure \ref{ocf-plot}, the oblate correction factors are depicted: the solid line is calculated by solving Equation (\ref{Fofd}) numerically, and the dashed line is obtained from our approximation. The excellent agreement between the approximated and the numeric $f_{\mathrm{OFD}}$ factors, at the pole and equator, demonstrates the robustness of this method. It is important to emphasize the intensity variation that results from the integration of  the $f_{\mathrm{OFD}}$ factor with co-latitude when gravity darkening effects are taken into account. In order to understand and compare the behavior of the scenario with gravity darkening in Figure \ref{oblado-factor} we also show the $f_{\mathrm{OFD}}$ without considering the gravity darkening. From the plots we can observe the large impact of the $f_{\mathrm{OFD}}$ on the polar direction, where the intensity between the models with and without gravity darkening follows a similar behaviour. Finally, with the purpose to test our calculation for the $f_{\mathrm{OFD}}$, we show in Figure \ref{ofd02} a comparison between the $f_{\mathrm{OFD}}$ with gravity darkening and the $f_{\mathrm{SFD}}$, both for a low value of $\Omega$ ($\Omega=0.20$) at the equator and the pole. At this rotational speed, the star remains almost spherical and the temperature essentially constant, confirming that both correction factors are similar.

\vskip 0.5cm
\section{Methodology for modelling a Be star disk}
\label{diskdens}
\subsection{Ad-hoc density model}
In this section we present a short summary of the traditional method to model a thin rotating circumstellar disk.
		 
An ad-hoc disk density model is generally adopted to describe the mass density stratification in the equatorial plane of a thin disk. This density distribution, originally developed by \citet{waters1987} for the interpretation of Be star IR observations, is assumed to follow a power-law distribution defined by:

\begin{equation}
\label{silaj}
\rho \left( r, z \right) = \rho_{0} \left(  \frac{r}{R_{*}} \right)^{-n} e^{-\left( z/H \right) ^{2}},
\end{equation}

\noindent where $\rho_{0}$ is the equatorial density of the disk at the stellar surface, $n$ is the adopted power-law index and $H$ is the scale height in the $z$-direction expressed by
\begin{equation}
H=\sqrt{\frac{2\, r^{3}}{\alpha_{0}}},
\end{equation}
\noindent where $\alpha_{0}$ is defined as 
\begin{equation}
\alpha_{0}= G\, M \, \frac{\mu_{0}\,m_{H}}{k\, T_{0}},
\end{equation}
\noindent where $\mu_{0}$ is the mean molecular weight of the gas and $T_{0}$ is an assumed isothermal temperature used for setting the scale height prior to the calculation of the self-consistent disk temperature distribution.

This simple prescription of a density distribution with power law fall off in the radial direction and in approximate hydrostatic equilibrium in the $z$ direction has been used extensively in the literature, so it provides a good basis for comparison with other works \citep[see, for example,][]{sigut2015,patel2016,arcos2017}.\\

Under the assumption of radiative equilibrium, the level populations and ionization state of the gas are calculated throughout the disk with the stellar radiation included using a Kurucz model atmosphere \citep{kurucz93}. In addition, the code solves the transfer equation along a set of rays parallel to the star's rotation axis. Then, by projecting the line flux at different angles, it is possible to calculate a synthetic line profile as seen by an external observer from a given line of sight  (see \cite{sigut2007} for details). 

\subsection{Hydrodynamic density model}

As mentioned in Section \ref{introd}, the spirit of this work is to extend the study developed by \cite{silaj2014b},  who modeled a Be star considering: i) a spherical star with constant effective temperature, and ii) a thin disk using density distributions provided by the $\Omega$-slow solution at the equatorial plane. They used these density distributions as input in the \textsc{Bedisk} code to obtain synthetic H$\alpha$ line profiles.
	
In this work we also include the effect of the stellar rotation on the shape of the star and gravitational darkening. The resulting radial density structure of the equatorial plane, obtained from the \textsc{Hydwind} code, is then used as input in the \textsc{Bedisk} code, to calculate the vertical densities ($z$-direction) and temperature distributions, and then the synthetic H$\alpha$ line profiles. It is important to note that the contribution from the outflowing wind at non-equatorial latitudes (fast wind solution) to the H$\alpha$ emission line profile is negligible, as explained in Section \ref{sumdis}. Thus, the H$\alpha$ emission line-forming region is primarily in or near the equatorial plane.

Here, it is worth mentioning that the radial density distribution obtained with the \textsc{Hydwind} code is calculated assuming conservation of angular momentum. 
However, based on observations, it is commonly accepted that the disks of Be stars are in Keplerian orbits, and therefore, in order to satisfy this observational feature, \textsc{Bedisk} computes the synthetic line profile with a Keplerian rotational distribution to obtain the emission line as a function of wavelength.
	
\section{Results}
\label{results}
	
In this work we solve an improved rotating radiation-driven wind model considering three different scenarios: 1) a spherically symmetric star with constant temperature, 2) an intermediate scenario that considers an oblate star but neglects gravity darkening effects, and 3) an oblate star with gravity darkening. For each scenario we calculate the density stratification from the EoM ($\Omega$-slow solution, for $\Omega > 0.75$) in the equatorial plane and obtain the corresponding synthetic H$\alpha$ line profiles from \textsc{Bedisk}. To investigate the results of these three scenarios, we first build a grid of hydrodynamical models.

\subsection{Grid of hydrodynamical models}
\label{grid_hyd}
We adopt the same stellar parameters as \cite{silaj2014b}: $T_{\mathrm{eff}}=25,000 \, \mathrm{K}$, $\log\, g = 4.03$,  $R_{*}=5.3 \, R_{\sun}$, which correspond to a B1V type star. Furthermore, we set $\rho_{*}(\pi/2) = 5 \times 10^{-11} \, \mathrm{g\, cm^{-3}}$ (see Equation \ref{denssup}) as a lower boundary condition for the \textsc{Hydwind} code. This boundary condition ensures that the initial surface velocity is less than the sound speed.

To represent a fast rotating star we use the following $\Omega$ values: $0.80$, $0.90$, $0.95$ and $0.99$. Then, the corresponding values for $\Gamma_{\Omega}$ are 0.026, 0.028, 0.030 and 0.032, which are slightly larger than $\Gamma_{\mathrm{E}}= 0.022$.

To date, there are no self-consistent values for the line-force parameters for the $\Omega$-slow solution. Therefore, we adopt values for  $\alpha$ and $\delta$ that are within the typical ranges calculated previously with both LTE \citep{abbott1982} and NLTE \citep{ppk1986} approximations, and we let the line-force parameter $k$ vary in a wider range. The line-force parameters used to construct the grid of models are shown in Table \ref{table1}. A total of 768 models were calculated initially with \textsc{Hydwind} for each of our three scenarios. 

Table~\ref{table2} shows the stellar parameters as a function of the rotational speed for the polar and equatorial directions for all three scenarios. The values for the equatorial region are taken directly from  Table~\ref{table2} for our calculations for the three different scenarios. Rotational velocities are obtained from the values of $\Omega$ and $v_{\mathrm{crit}}$ and, according to Equations (\ref{vcrit1d}), (\ref{vcrit2d}) and (\ref{vcrit}), we derive $v_{crit} = 622\,\mathrm{km\, s^{-1}}$, $508\,\mathrm{km\, s^{-1}}$ and $513\,\mathrm{km\, s^{-1}}$ for the spherical, oblate, and oblate plus gravity darkening scenarios, respectively.

It is worth noting that the \textsc{Hydwind} code can only obtain hydrodynamic solutions for some combinations of the parameters of our grid, because not all combinations have physical stationary solutions \citep{venero2016}. 

\subsection{Density distributions}
\label{density}

\begin{figure}[h!]
	\center
	\includegraphics[width=3 in]{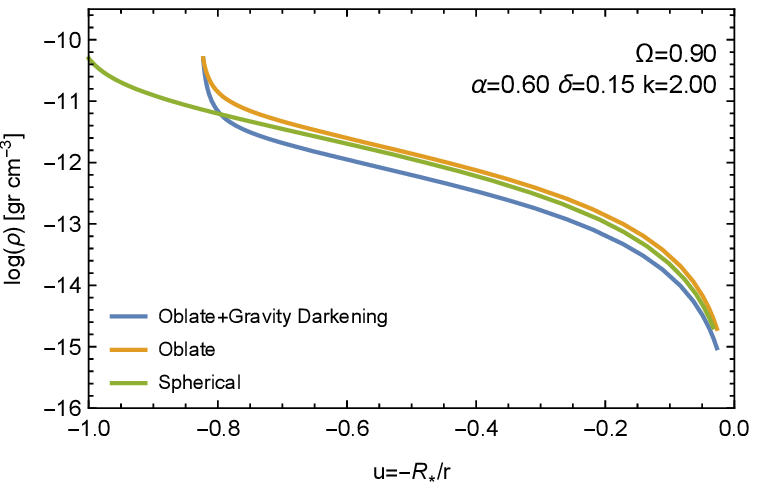}\\
	\vskip 0.3cm		
	\includegraphics[width=3 in]{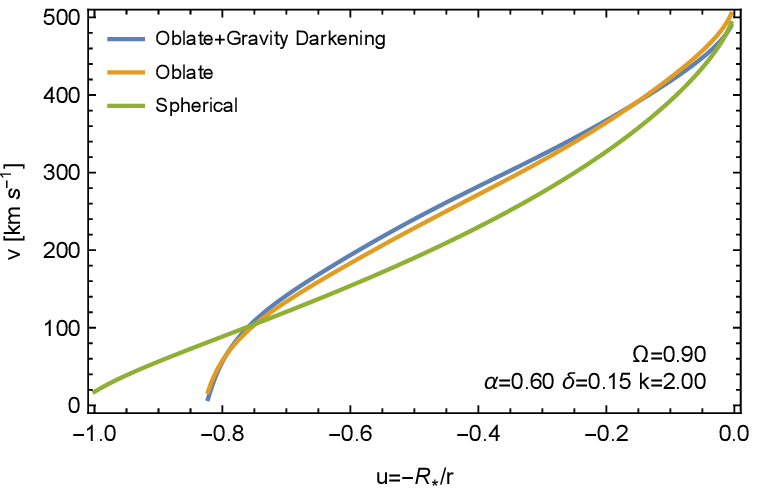}
	\caption{Mass density distributions  (upper panel) and velocity profiles (lower panel) at the equatorial direction computed for rotating radiation-driven winds  with $\Omega=0.90$ for the three scenarios as function of the inverse radial coordinate $u$.}  \label{hydrocomp}
\end{figure}

Figure \ref{hydrocomp} shows the density distribution (upper panel) and velocity profiles (lower panel) at the equatorial plane for hydrodynamic models with $\Omega=0.90$ for our three scenarios. These models show similar terminal velocities, $v_{\infty}=$ $493, 505$, and $489$ $\mathrm{km\, s^{-1}}$, respectively. Although the models presented in  Figure \ref{hydrocomp} have the same set of line-force parameters, their mass-loss rates are different, namely $\dot{M}= 2.36 \times 10^{-6},\, 3.46 \times 10^{-6}$, and $1.63 \times 10^{-6}\,$ M$_\odot$ yr$^{-1}$. Both oblate models show very similar velocity profiles, but the mass-loss rate from scenario 3 is about half the value of scenario 2. This is due to the decrease in effective temperature with latitude that mainly affects the calculation of the $f_{\mathrm{OFD}}$ factor and the radiative flux coming from the star. The difference is seen in the density distribution. Notice that in the oblate scenarios, the stellar surface in the $u$ coordinate starts at $u>-1$  due to the fact that the equatorial radius is larger than the polar one.
	
\begin{figure}[h!]
	\center
	\includegraphics[width=3 in]{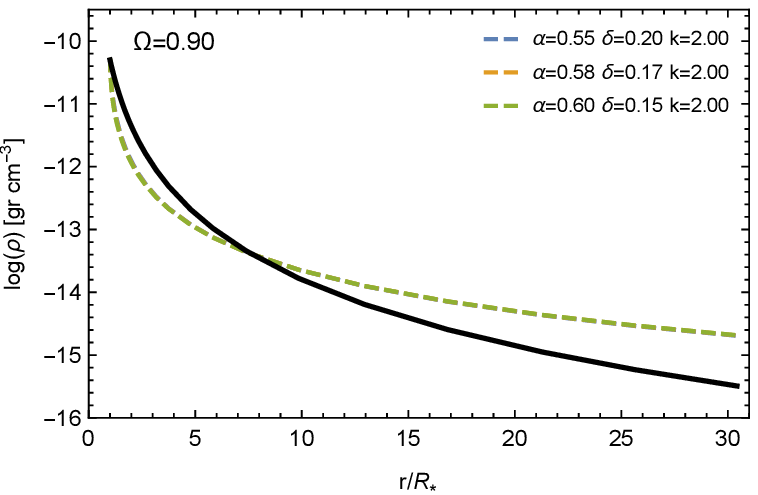}\\
	\vskip 0.3cm	
	\includegraphics[width=3 in]{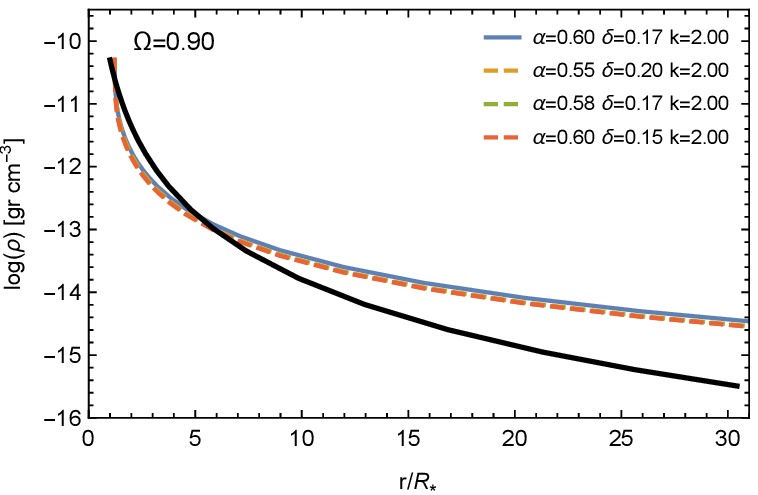}\\
	\vskip 0.3cm	
	\includegraphics[width=3 in]{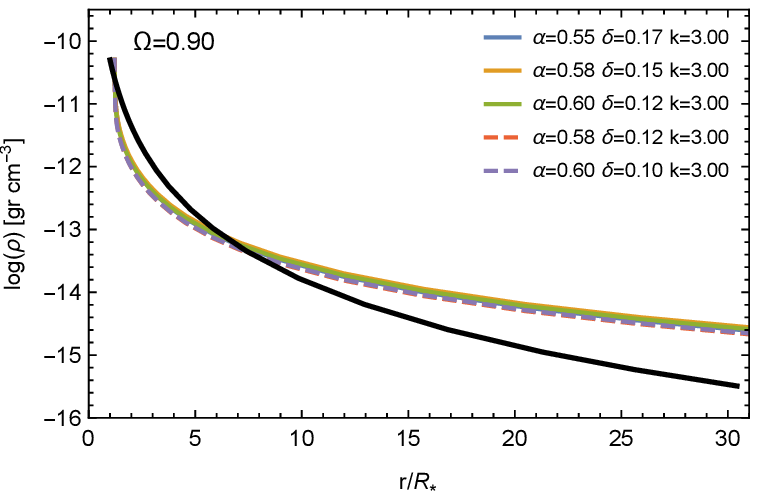}
	\caption{Mass density distributions as a function of $r/R_{*}$ for the equatorial direction computed for rotating radiation-driven winds with $\Omega=0.90$ and  different sets of line-force parameters.   The equatorial mass density distributions  are compared with the ad-hoc mass density structure (solid black line). These models correspond to the best fit line profile to the ad-hoc model, using the finite disk correction factor for: i) a spherical symmetric star (top panel), ii)  an oblate star (middle panel), and iii) an oblate star with gravity darkening (bottom panel).}   \label{P-1d}
\end{figure}

Figure \ref{P-1d} compares some equatorial mass density distributions from our grid with the ad-hoc mass density structure (solid black line). This ad-hoc density model (Equation \ref{silaj}) is calculated with $\rho_0 = \rho_{*}(\pi/2)$ and $n=3.5$ as in \citet{silaj2014b}. For all the hydrodynamical cases, the mass density shows almost the same characteristic behaviour. Near the surface of the star, the hydrodynamic mass density structures fall faster than the ad-hoc model up to 5 to 8 stellar radii, and at greater distances, the density from the ad-hoc model decays faster.

\subsection{Synthetic H$\alpha$ line profiles}
\label{profiles}

All mass density distributions resulting from our computations for the different scenarios were used to compute a grid  of H$\alpha$ line profiles with the \textsc{Bedisk} code.

\begin{figure}[h]
	\center
	\includegraphics[width=3 in]{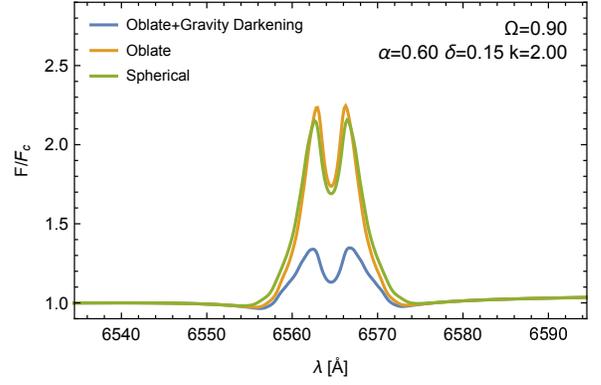}
	\caption{Comparison of H$\alpha$ line profiles obtained from the \textsc{Bedisk} code using the density distributions shown in Figure \ref{hydrocomp}. The line profile intensities for each scenario are due to the different mass-loss rates.  \label{SOGD}} 
\end{figure}

Figure \ref{SOGD} shows the resulting H$\alpha$ line profiles from the different mass density distributions shown in Figure \ref{hydrocomp}. Since the H$\alpha$ emission line is very sensitive to the mass density distribution ($\propto\, \rho^2$), higher densities correspond to stronger emission. As a consequence, the scenario including the gravity darkening effect has the lowest intensity in the emission line profile, because it has a lower mass-loss rate. 
		
To select H$\alpha$ profiles from our grid that best-fit the profile obtained from the ad-hoc density model, we define the following selection criteria: 
a) our computed line profile is considered similar to the ad--hoc synthetic H$\alpha$ profile when the discrepancy between the maximum intensities of the line emission of both models is lower than  a $15\%$, and b) we adopt the smallest value of the $k$ parameter to ensure that the $k$ values remain physically reasonable. 

\begin{figure}[h]
	\center
	\includegraphics[width=3 in]{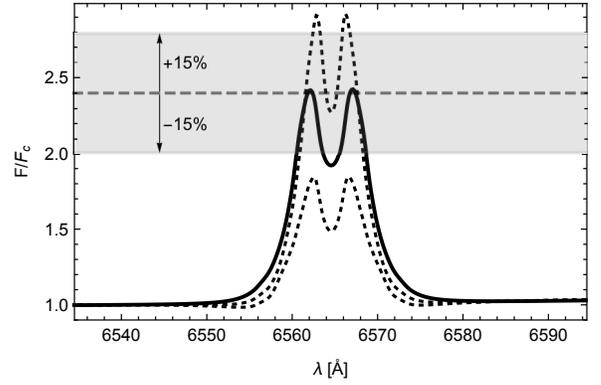}
	\caption{Comparison of the ad-hoc H$\alpha$ line profile (solid line) with two models (dotted lines) that lie outside our criteria (shadowed region), i. e., a discrepancy of $\pm15\%$ between the maximum intensity of the ad-hoc line profile (horizontal dashed line) and the maximum intensity of the profile from a calculated model. \label{figquant}}
\end{figure}

Figure \ref{figquant} shows the ad-hoc H$\alpha$ line profile compared with two profiles that are not selected by this criteria.

Table \ref{table3} summarizes the model parameters of the best fitting synthetic H$\alpha$ profiles. Our results for each scenario are: 

\textit{a) Star with Spherical Symmetry:}
For this case, we find that the best models have a value of $k=3.0$ at $\Omega=0.80$, and $k=2.0$ for the higher rotation rates. The $\alpha$ parameter varies from 0.55 to 0.60 with the exception of the model with $\Omega=0.80$ where $\alpha=0.50$. The $\delta$ parameter ranges from 0.15 to 0.20, except for the case with $\Omega=0.99$ which has $\delta=0.12$.\\

\textit{b) Oblate Star:}
The best models have values of $k=2.0$  for $\Omega=0.80$ and $\Omega=0.90$, and values of $k$ between 1.5 and 2.0 for higher values of $\Omega$. The $\alpha$ parameter has values from $0.55$ to $0.60$ with the exception of one model with $\alpha=0.50$ at $\Omega=0.99$. The $\delta$ parameters, for this scenario, have a range from 0.15 to 0.20, except for a case with $\Omega=0.99$ which has $\delta=0.12$.\\

\textit{c) Oblate Star with Gravity Darkening:}
For this scenario, we were not able to obtain appreciable line emission when  $\Omega=0.99$. For the other rotation rates, even for models with similar $\dot{M}$ as in previous scenarios, the best models have larger values of $k$ ($k=3.0$) as a consequence of a reduction of the radiation flux. The $\alpha$ parameters are between 0.55 and 0.60, and the $\delta$ parameters range from 0.12 to 0.20.\\

\begin{figure}[h]
\center
\includegraphics[width=3 in]{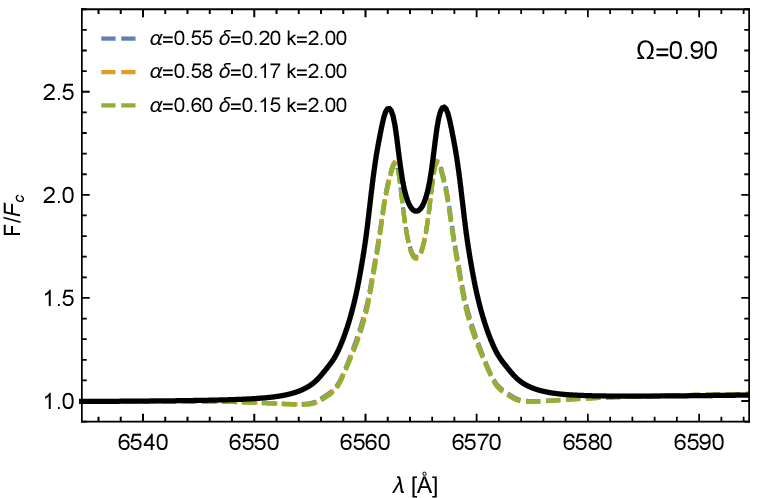}\\
\vskip 0.3cm	
\includegraphics[width=3 in]{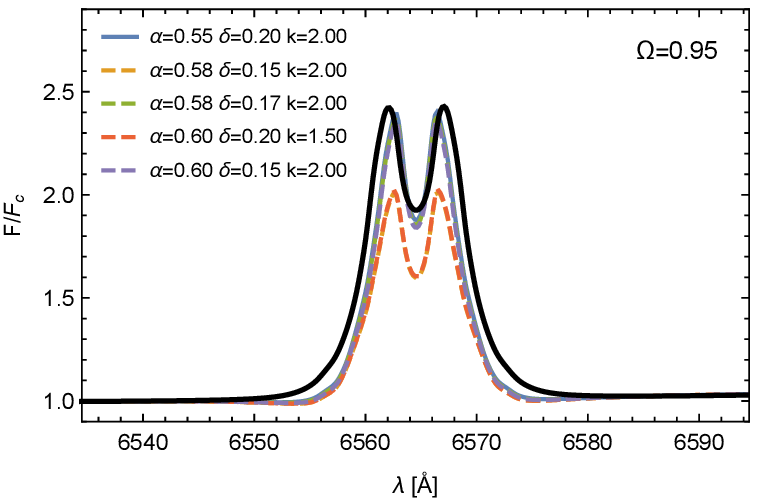}
\caption{Synthetic H$\alpha$ profiles computed at $\Omega=0.90$ and $0.95$ with the radiative transfer code \textsc{Bedisk}, using the equatorial density structure obtained from the hydrodynamic code \textsc{Hydwind}, assuming spherical symmetry for the star. The adopted rotational velocity is given in the top right hand corner of each panel, and the line-force parameters used to calculate the density structure are given in the legend. Dashed and solid lines correspond to profiles with a lower and higher intensity than the ad-hoc profile, respectively. The thick black line corresponds to the emission line profile computed with the ad-hoc density structure. An inclination of $i=35 \degree$ is assumed for all profiles.  \label{Ha-1d-S}} 
\end{figure}

\begin{figure}[h]
\center
\includegraphics[width=3 in]{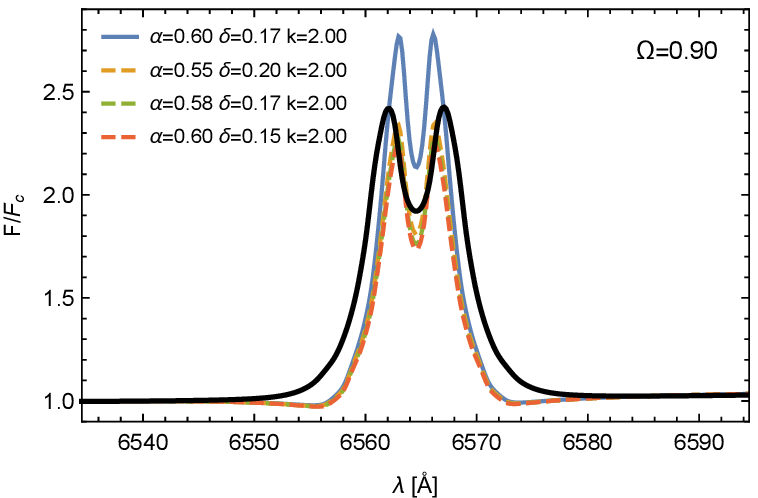}\\
\vskip 0.3cm	
\includegraphics[width=3 in]{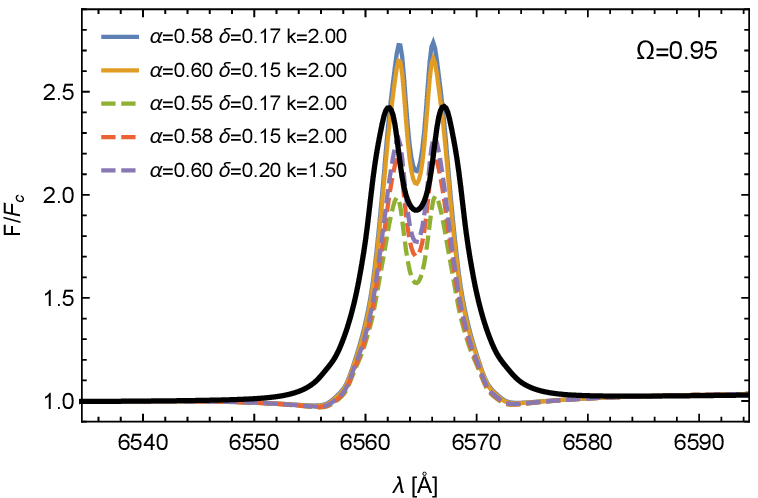}
\caption{Similar to Figure \ref{Ha-1d-S}, but with the equatorial density structure obtained calculated assuming an oblate shape for the star and neglecting the gravity darkening effect. \label{Ha-1d-O}} 
\end{figure}

\begin{figure}
\center
\includegraphics[width=3 in]{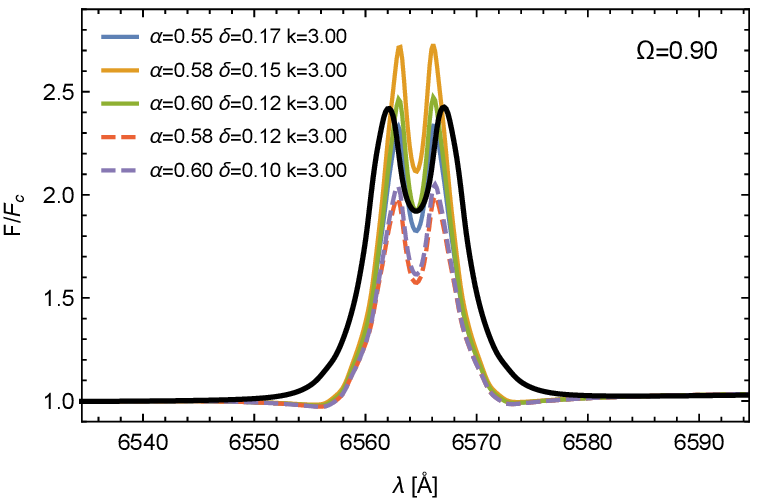}\\
\vskip 0.3cm	
\includegraphics[width=3 in]{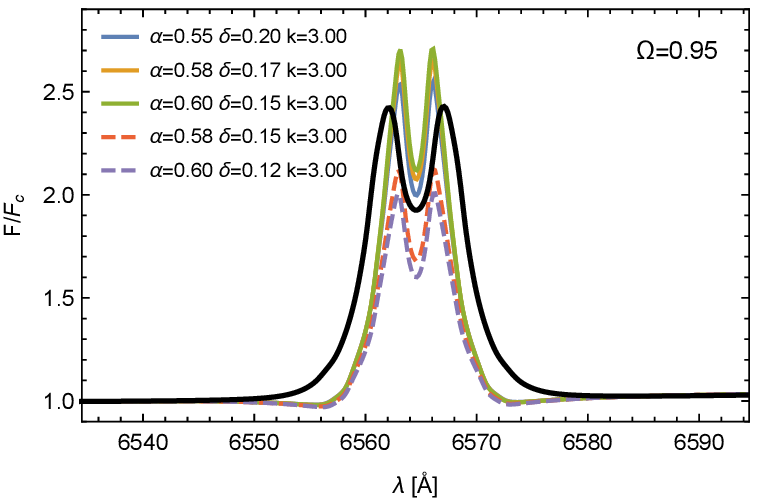}
\caption{Similar to Figure \ref{Ha-1d-S}, but with the equatorial density structure calculated assuming an oblate shape for the star and including the gravity darkening effect. \label{Ha-1d-GD}} 
\end{figure}

The synthetic H$\alpha$ profiles that correspond to the best hydrodynamical models (selected by our criteria) with $\Omega=0.90$ and $\Omega=0.95$  are depicted in Figure \ref{Ha-1d-S}, \ref{Ha-1d-O}, and \ref{Ha-1d-GD} for each scenario, respectively. The solid black line corresponds to the emission line profile computed with the ad-hoc density structure and the dashed lines represent synthetic profiles from the hydrodynamical models. All profiles are calculated for an observer's line of sight to the star of $i=35 \degree$.

In general, for our three scenarios, a combination of the highest values of $\alpha$ and $\delta$ from our grid is necessary to obtain an emission profile (with a low $k$) similar to our ad-hoc model. The impact of $\alpha$ and $\delta$ on the wind parameters ($\dot{M}$, $v_{\infty}$) corresponding to the $\Omega$-slow solution is similar to the behavior expected from the fast solution: smaller values of $\alpha$ have correspondingly smaller $\dot{M}$ and $v_{\infty}$, while with smaller $\delta$, $\dot{M}$ is smaller and $v_{\infty}$ is higher. For a given $k$, note that   the $\alpha$ value has a greater impact on the final wind solution than the $\delta$ value.

Finally, as our hydrodynamical models have lower density values than the ad-hoc model near the surface of the star (see Figure \ref{P-1d}), the emission in the line wings of our model does not appear as strong as it does in the ad-hoc models.

\section{Summary and Discussion \label{sumdis}}

We implemented the effect of high stellar rotation on line-driven winds by adopting a non-spherical central star, and applying gravity darkening and the oblate finite disk correction factor to the m-CAK model. In order to  numerically solve this improved 
model we developed an iterative procedure to calculate the oblate finite disk correction factor. This is a more general and robust model than the one developed by \citet{muevink14}, because it retains the topology of the spherical m-CAK model, i.e., the model can be used to obtain either fast solutions \citep{friend1986}, $\Omega$-slow solutions \citep{cure2004}, or $\delta$-slow solutions \citep{cure2011,venero2016}. When $\Omega \gtrsim 0.75$, our wind model describes a two-component wind regime similar to that obtained by \cite{cure2005}. 

From our results, we highlight the role of the oblate finite disk correction factor in the hydrodynamic,
velocity, and density profiles, when these are compared with the spherical cases. In the case of the oblate finite disk correction factor calculated with gravity darkening, $f_{\mathrm{OFD}}(u)$ varies by at least a factor of 2 between the equator and pole. Moreover, this particular derivation of $f_{\mathrm{OFD}}(u)$ results in a decrease in the mass-loss rate when compared with the factor obtained from the oblate case without gravity darkening. This is due to 1) the decrease of the temperature and 2) with increasing distance from the star, larger stellar surfaces with different temperatures than the radial only direction, contribute.

Next we investigated the influence of oblateness and gravity darkening effects on the
emergent H$\alpha$ line profile from a Be star circumstellar disk. For this, we used the equatorial mass density distribution as input in the \textsc{Bedisk} code, neglecting the effects of radial and azimuthal velocities from the hydrodynamics in the radiative transport calculations.

We created a grid of equatorial $\Omega$-slow models and the resulting grid of H$\alpha$ profiles, for three different scenarios: 1) a spherically symmetric star with constant temperature, 2) an oblate star with constant temperature, and 3) an oblate star with gravity darkening.
The results from the H$\alpha$ grid were compared with the ad-hoc line profile, that follows a $r^{-n}$  mass density distribution with $n=3.5$ and $\rho_{0}=5.0 \times 10^{-11}\, \mathrm{g \, cm^{-3}}$.

\begin{figure}[h]
\center 
\includegraphics[width=3 in]{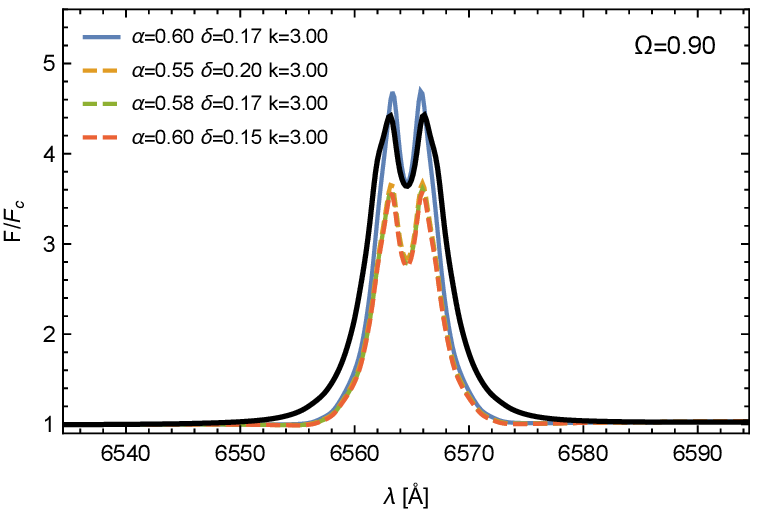}\\
\vskip 0.3cm	
\includegraphics[width=3 in]{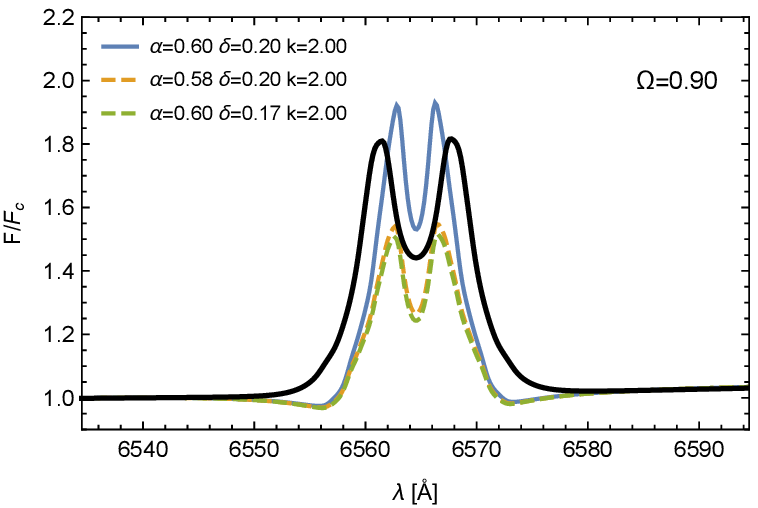}
\caption{H$\alpha$ profiles: the ad-hoc model for $n=3$ is shown in the solid black line (upper panel). Other profiles from our grid of hydrodynamic models that satisfy our selection criteria are shown in dashed lines (see legend). Bottom panel is same as the upper panel, but for $n=4$. See text for details.}\label{general}
\end{figure}

We found an agreement between hydrodynamical and ad-hoc H$\alpha$ emission profiles (matching our 15\% criteria)  for values of $k \sim 1.5\,$to $\,2$ for spherical or oblate cases, and $k \sim 3$ for the oblate case with gravity darkening, and values of $\alpha \sim  0.6$ and $\delta \gtrsim 0.1$. These results are for an ad-hoc model with $n=3.5$. In order to explore the variation in the line-force parameters and match ad-hoc models with n=3 (stronger emission line)  and n=4 (weaker emission line), we selected models based on our selection criteria, explained previously. We obtain good agreement between the oblate hydrodynamical models with gravity darkening and ad-hoc models for values of $k \gtrsim 3$  when $n=3$ and $k \gtrsim 2$ when $n=4$ (see Figure \ref{general}). Models with $n=3$ have slightly higher values of $\alpha$ and $\delta$ than models with $n=3.5$.
		
To date, even though there is not a self-consistent calculation of the line-force parameters for the $\Omega$-slow solution, for our scenarios 1 and 2, the values of $k$ are roughly of order $1/\alpha$, which, according to \citet{puls2000}, would be a maximum value if line-overlap effects are neglected. We suggest, for scenario 3, that the large values found for $k$  may be related to changes in the ionization structure and possible increments in the opacity of the absorption lines along the disk together with line-overlap effects. 


On the other hand, to reproduce the emission line from an ad-hoc model with lower values of the exponent $n$, we require higher values of $k$, hence multi-line scattering processes should be considered for these dense disks. Nevertheless, the values of $k$ obtained in this study are closer to predicted values than those obtained by \cite{silaj2014b}. Finally, it is worth noting that for a given $k$ the effect of $\alpha$ in the final wind solution is stronger than the effect of $\delta$, generally. 


Regarding the effect of the rotational velocity on the mass-loss rate, it is the corresponding equatorial mass density that mainly determines the strength of the emission profiles. However, for the same set of line-force parameters ($\alpha$, $k$, $\delta$), higher $\Omega$ values produce greater intensity of the emission line; this is valid for the scenarios assuming a spherical and an oblate star of constant temperature. For the third scenario (an oblate star with gravity darkening), the maximum line intensity is attained at a value of $\Omega < 1$. When $\Omega$ tends toward one, the emission is lower due to the decrease in effective temperature because of gravity darkening.

\begin{figure}[h]
\center 
\includegraphics[width=3 in]{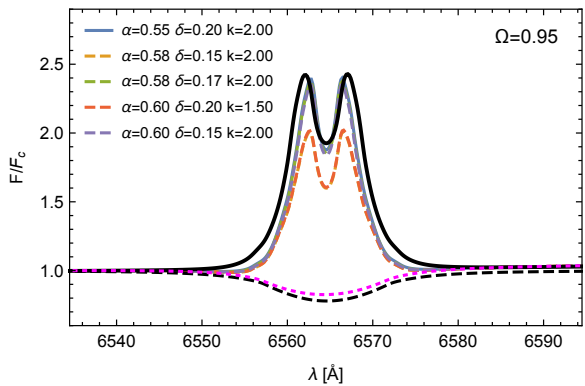}
\caption{H$\alpha$ profiles: a) absorption profile obtained for the non-disk region of a spherical rotating model at $\Omega=0.95$, computed using the \textsc{Fastwind} code (black dashed line); b) absorption profile of the stellar photosphere calculated by \textsc{Bedisk} (magenta dotted-line); c) emission profiles obtained for the disk plus photosphere calculated by \textsc{Bedisk} (see details in Figure \ref{Ha-1d-S}). Note that the absorption profile calculated by \textsc{Fastwind} from the non-disk region includes the photospheric contribution, demonstrating that the overall contribution from the wind of the non-disk region is minimal.}\label{fw}
\end{figure}

The contribution of the fast wind component is expected to have a negligible effect on the emission profile of  the H$\alpha$ line due to the low density of this region. In order to test this argument, we calculated the contribution to H$\alpha$ using a fast wind solution with the following line-force parameters: $\alpha$ = 0.565, $k$ = 0.32 and $\delta$ = 0.02, based on the calculations from \cite{ppk1986} for a model with a similar $T_{\rm{eff}}$ as our star. From the hydrodynamic calculations with \textsc{Hydwind} for a spherical non-rotating case, a mass-loss rate of 1.2 $\times$ 10$^{-8}$ M$_\odot$ yr$^{-1}$ and a terminal velocity of 2350 km\,s$^{-1}$ are obtained. Then, this hydrodynamic solution is used as input in the radiative transfer code \textsc{Fastwind} \citep{puls2005} with the aim of obtaining the line profile accounting for the total contribution from the stellar photosphere and the fast wind (at all latitudes). We obtain an absorption line profile which is then convolved with $v\,\sin(35\degree)$ = 339 km\,s$^{-1}$. This profile is shown in Figure \ref{fw} in the black dashed line. Furthermore, this figure also shows the contribution to the H$\alpha$ line from the stellar photosphere (absorption profile in magenta dashed line) calculated by \textsc{Bedisk} and  the emergent emission profiles from the stellar photosphere plus the disk (same as lower panel of Figure \ref{Ha-1d-S}). We note that the small difference between the absorption profiles demonstrates that, as expected, there is a minimal (neglectable) contribution from the fast wind component.

 Finally, we have shown that the mass density distribution obtained from the $\Omega$-slow wind solution (with $k\,\sim$ 3), considering a more realistic scenario with gravity darkening,  is able to reproduce  H$\alpha$ line emission features similar to those predicted by a power-law model. In addition, we also require a mechanism to provide Keplerian rotation within the disks of our models to produce reasonable profile shapes, which is an area of ongoing research in the field of Be stars.

In view of these new results, we are encouraged to further develop this line of research. In a future work, we plan to calculate, in a self-consistent way, the line force parameters of this $\Omega$-slow solution. In addition, it would be interesting to analyse the stability of this two-component wind in a 2D time dependent frame that includes non-radial forces.

\begin{acknowledgements}
The authors would like to thank the referee, Dr. Joachim Puls, for his thoughtful comments and suggestions that helped improve the paper significantly. This research was supported by the Canada-Chile Leadership Exchange Scholarship program from Government of Canada. I. A. acknowledges support from Fondo Institucional de Becas FIB-UV and Gemini-Conicyt 32120033. C. E. J. thanks support from NSERC, the National Sciences and Engineering Research Council of Canada. M. C. acknowledges support from Centro de Astrof\'isica de Valpara\'iso. 
L. C. and M. C. thank support from the project CONICYT + PAI/Atracci\'on de capital humano avanzado del extranjero (folio PAI80160057). L. C. also acknowledges financial support from Universidad Nacional de La Plata (Programa de incentivos 11/G137) and form CONICET (PIP 0177).
A. G. acknowledges support from the Swiss National Science Foundation through the Advanced Postdoc Mobility fellowship, project P300P2$\_$158443. A. J. acknowledges support from CONICYT FONDECYT/POSTDOCTORADO 3150673, Nucleo Milenio ICR RC130003 and Proyecto Anillo ACT 1112. 
\end{acknowledgements}

\software{\textsc{Hydwind} \citep{cure2004},
               \textsc{Bedisk} \citep{sigut2007}, 
               \textsc{Fastwind} \citep{puls2005}
               }

\clearpage
\begin{deluxetable}{cccccccccc}
\tablecolumns{10}
\tabletypesize{\normalsize}
\tablecaption{Combinations of the line--force parameters  for the grid of models. \label{table1}}
\tablewidth{0pt}
\tablehead{\colhead{Parameter} & \multicolumn{8}{c}{Values}}
\startdata
$\alpha$ & 0.50 & 0.55 & 0.58 & 0.60 \\
$\delta$ & 0.07 & 0.10 & 0.12 & 0.15 & 0.17 & 0.20 \\
$k$        & 0.50 & 0.80 & 1.00 & 1.50 & 2.00 & 3.00 & 4.00 & 5.00 \\
\enddata
\end{deluxetable}

\clearpage
\begin{deluxetable}{cccccccccccc}
\tablecolumns{6}
\tabletypesize{\normalsize}
\tablecaption{Stellar parameters used in our calculations for the three different scenarios. \label{table2}}
\tablewidth{0pt}
\tablehead{\colhead{$\Omega$} & \colhead{$R_{\mathrm{eq}}$} & \colhead{$T_{\mathrm{eq}}$} & \colhead{$T_{\mathrm{pole}}$} & \colhead{$v_{\mathrm{rot}}(\mathrm{eq})$} & \colhead{Scenario}\\
\colhead{} & \colhead{$[R_{\odot}]$} & \colhead{[K]} & \colhead{[K]} & \colhead{$[\mathrm{km\, s^{-1}}]$} & \colhead{}}
\startdata
0.80 & 5.30 & 25 000 & 25 000 & 497.5 & Spherical \\
        & 6.05 & 25 000 & 25 000 & 308.9 & Oblate \\
        & 6.05 & 22 770 & 25 802 & 312.2 & Oblate + Gravity Darkening\\
\tableline        
0.90 & 5.30 & 25 000 & 25 000 & 559.6 & Spherical \\
        & 6.44 & 25 000 & 25 000 & 370.2 & Oblate \\
        & 6.44 & 21 635 & 26 020 & 374.3 & Oblate + Gravity Darkening\\ 
\tableline     
0.95 & 5.30 & 25 000 & 25 000 & 590.7 & Spherical \\
        & 6.79 & 25 000 & 25 000 & 411.9 & Oblate \\
        & 6.79 & 20 617 & 26 139 & 416.4 & Oblate + Gravity Darkening\\
\tableline             
0.99 & 5.30 & 25 000 & 25 000 & 615.5 & Spherical \\
        & 7.37 & 25 000 & 25 000 & 465.6 & Oblate \\
        & 7.37 & 18 698 & 26 240 & 470.7 & Oblate + Gravity Darkening\\                        
\enddata
\end{deluxetable}

\begin{deluxetable*}{ccccccccccccc}
\tablecolumns{12}
\tabletypesize{\normalsize}
\tablecaption{Wind and line--force parameters for $\Omega= 0.8, 0.9, 0.95$ and $0.99$, showing synthetic H$\alpha$ profiles similar to the ad-hoc profile. The models are computed with an equatorial density structure. Terminal velocities are calculated at 100 stellar radii. \label{table3}}
\tablewidth{0pt}
\tablehead{\colhead{} &\colhead{} &\colhead{} &\colhead{} & \multicolumn{2}{c}{Spherical Shape}&\colhead{}& \multicolumn{2}{c}{Oblate Shape}&\colhead{}& \multicolumn{2}{c}{Oblate Shape + Gravity Darkening}\\
\cline{5-6} \cline{8-9} \cline{11-12}\\
\colhead{$\Omega$} & \colhead{$\alpha$} & \colhead{$\delta$} & \colhead{$k$} & \colhead{$\dot{M}\left( \theta=90\degree  \right) $} & \colhead{$v_{\infty}\left( \theta=90\degree  \right)$}&\colhead{}& \colhead{$\dot{M}\left( \theta=90\degree  \right)$} & \colhead{$v_{\infty}\left( \theta=90\degree  \right)$}&\colhead{}& \colhead{$\dot{M}\left( \theta=90\degree  \right)$} & \colhead{$v_{\infty}\left( \theta=90\degree  \right)$}\\
\colhead{}&\colhead{} &\colhead{} &\colhead{} &\colhead{$[10^{-6}\, \mathrm{M_{\odot}\, yr^{-1}}]$} & \colhead{$[\mathrm{km\, s^{-1}}]$}&\colhead{}&\colhead{$[10^{-6}\, \mathrm{M_{\odot}\, yr^{-1}}]$} & \colhead{$[\mathrm{km\, s^{-1}}]$} &\colhead{}& \colhead{$[10^{-6}\, \mathrm{M_{\odot}\, yr^{-1}}]$} & \colhead{$[\mathrm{km\, s^{-1}}]$}
}
\startdata
0.80    & 0.50 & 0.17 & 3.00 & 2.04 & 421.22  & & \nodata  & \nodata   & & \nodata & \nodata \\
        & 0.55 & 0.12 & 3.00 & 2.44 & 494.07  & & \nodata  & \nodata   & & \nodata & \nodata \\
        & 0.55 & 0.17 & 3.00 & \nodata & \nodata  & & \nodata  & \nodata   & & 3.17 & 496.51 \\
        & 0.58 & 0.15 & 3.00 & \nodata & \nodata  & & \nodata  & \nodata   & & 3.80 & 534.54 \\
        & 0.58 & 0.20 & 2.00 & \nodata & \nodata  & & 3.69  &  526.29  & & \nodata & \nodata \\
        & 0.60 & 0.17 & 2.00 & \nodata & \nodata  & & 3.49  &  558.34  & & \nodata & \nodata \\
\tableline     
0.90    & 0.55 & 0.17 & 3.00 & \nodata & \nodata  & &  \nodata & \nodata & &  2.61     & 438.60   \\
        & 0.55 & 0.20 & 2.00 & 1.95      & 419.93  & &   3.10   & 443.01  & &  \nodata  & \nodata \\
        & 0.58 & 0.12 & 3.00 & \nodata & \nodata  & &  \nodata & \nodata & &  2.53     & 482.20   \\
        & 0.58 & 0.15 & 3.00 & \nodata & \nodata  & &  \nodata & \nodata & &  3.20     & 471.83   \\
        & 0.58 & 0.17 & 2.00 & 2.19      & 462.54  & &   3.31   & 479.51  & &  \nodata  & \nodata \\
        & 0.60 & 0.10 & 3.00 & \nodata & \nodata  & &  \nodata & \nodata & &  2.76     & 507.16   \\
        & 0.60 & 0.12 & 3.00 & \nodata & \nodata  & &  \nodata & \nodata & &  3.18     & 500.36   \\
        & 0.60 & 0.15 & 2.00 & 2.36      & 492.80  & &   3.46   & 504.56  & &  \nodata  & \nodata \\
        & 0.60 & 0.17 & 2.00 & \nodata & \nodata  & &  4.17     & 498.29  & & \nodata  & \nodata  \\
%
%
\tableline      	  
0.95   	 & 0.55 & 0.17 & 2.00 & \nodata & \nodata  & & 2.60      & 419.66  & &  \nodata  & \nodata  \\
         & 0.55 & 0.20 & 2.00 & 2.02    & 399.08   & &  \nodata  & \nodata & &  \nodata  & \nodata \\
     	 & 0.55 & 0.20 & 3.00 & \nodata & \nodata  & &  \nodata  & \nodata  & &  2.58     & 401.07  \\
     	 & 0.58 & 0.15 & 2.00 & 1.88      & 447.03  & &  3.11    & 450.64  & & \nodata & \nodata \\
     	 & 0.58 & 0.15 & 3.00 & \nodata & \nodata  & &  \nodata & \nodata & &  2.49     & 442.28  \\
     	 & 0.58 & 0.17 & 2.00 & 2.25      & 439.93  & &  3.78    & 445.10  & & \nodata & \nodata \\
     	 & 0.58 & 0.17 & 3.00 & \nodata & \nodata  & &  \nodata & \nodata & &  2.93     & 435.55  \\
     	 & 0.60 & 0.12 & 3.00 & \nodata & \nodata  & &  \nodata & \nodata & &  2.55     & 469.50  \\
     	 & 0.60 & 0.15 & 2.00 & 2.41      & 469.05  & &  3.89    & 468.08  & & \nodata & \nodata \\
     	 & 0.60 & 0.15 & 3.00 & \nodata & \nodata  & &  \nodata & \nodata & &  3.17     & 459.28  \\
     	 & 0.60 & 0.20 & 1.50 & 1.89      & 450.19  & &  3.22    & 452.46  & & \nodata & \nodata \\
%
%
\tableline       	 
0.99  	 & 0.50 & 0.20 & 2.00 & \nodata & \nodata  & &  2.17      & 340.64  & & \nodata & \nodata \\
     	 & 0.55 & 0.17 & 2.00 & \nodata & \nodata  & &  3.11      & 385.51  & & \nodata & \nodata \\
     	 & 0.58 & 0.15 & 2.00 & \nodata & \nodata  & &  3.62      & 413.84  & & \nodata & \nodata \\
         & 0.58 & 0.20 & 1.50 & \nodata & \nodata  & &  2.99      & 399.77  & & \nodata & \nodata \\
         & 0.60 & 0.12 & 2.00 & 1.93    & 462.09   & &  3.42      & 437.32  & & \nodata & \nodata \\
     	 & 0.60 & 0.17 & 1.50 & \nodata & \nodata  & &  2.84      & 423.22  & & \nodata & \nodata \\
%
\enddata
\end{deluxetable*}
\clearpage
\appendix

\section{Equation of Motion for an oblate gravity darkened star}
\label{eq-GD}
To derive the Equation of Motion to account for gravity darkening and the oblate distortion of the star in the radial direction, we start by analyzing the total (radial) acceleration (sum of effective and radiative accelerations) on the barotropic stellar layers, the photosphere. Following the work of \cite{maeder2000a} we have 

\begin{equation}
\label{mmgtot}
g_{\mathrm{tot}}=g_{\mathrm{eff}}+g_{\mathrm{rad}} \approx g_{\mathrm{eff}}(1-\Gamma_{\Omega}), 
\end{equation}

\noindent where we have evaluated $\Gamma_{\Omega}$ with $\sigma_{\mathrm{E}}$. This is a reasonable approximation within the following context, due to the difficulty of knowing the exact value of $\kappa$, and because the upper photosphere of hot stars is dominated by electron scattering and to a lesser degree by line processes. As the line contribution is small below the sonic point we can generalize \citet{maeder2000a}'s expression for the wind region by adding an appropriate approximation for the line acceleration term that takes into account the increasing line force due to Doppler effects, allowing $g_{\mathrm{tot}}>0$, i. e., 

\begin{equation}
\label{gmaeder}
g_{\mathrm{tot}} \approx g_{\mathrm{eff}}(1-\Gamma_{\Omega}) + g_{\mathrm{rad}}^{\mathrm{line}}.
\end{equation}

In the m-CAK formalism, the radiative line acceleration can be expressed in terms of the continuum radiative acceleration due to Thomson scattering times a multiplication factor $M(t)$, the so-called force-multiplier that is a function of the optical depth parameter $t$,

\begin{equation}
g_{\mathrm{rad}}^{\mathrm{line}} \equiv g_{\mathrm{rad}}^{\mathrm{Th}} \, M(t)
\end{equation}
 
\noindent with  

\begin{equation}
M(t) \equiv M_{\mathrm{point}}(t) \, f_{\mathrm{FD}},
\end{equation}

\noindent where $M_{\mathrm{point}}(t)$ is the force-multiplier when the star is assumed to be a point source and $f_{\mathrm{FD}}$ is the finite disk correction factor for a spherical or oblate star \citep[see, e.g.,][]{lamers1999, pelupessy2000}. Then, considering the expression 

\begin{equation}
\label{gth}
g_{\mathrm{rad}}^{\mathrm{Th}} \equiv \frac{\sigma_{\mathrm{E}}\, \mathcal{F}}{c} = \Gamma_{\mathrm{E}} \, g_{\mathrm{grav}},
\end{equation}

\noindent we have

\begin{equation}
g_{\mathrm{rad}}^{\mathrm{line}} =   \Gamma_{\mathrm{E}} \, g_{\mathrm{grav}} \, M_{\mathrm{point}}(t) \, f_{\mathrm{FD}},
\end{equation}

\noindent that, in turn, eventually can be generalized to consider gravity darkening if the local radiative flux and the appropriate finite disk correction factor is used. Thus, the radiative line acceleration can be expressed as

\begin{equation}
g_{\mathrm{rad}}^{\mathrm{line}}(\theta) =   \Gamma_{\mathrm{E}}(\theta) \, g_{\mathrm{grav}} \, M_{\mathrm{point}}(t) \, f_{\mathrm{OFD}},
\end{equation}

\noindent where $\Gamma_{\mathrm{E}}(\theta) $ is in function of the local flux. Considering this expression, then Equation \ref{gmaeder} can be written in the following way, 

\begin{equation}
\label{eq-general}
g_{\mathrm{tot}}(\theta) \approx -g_{\mathrm{grav}}(1-\Gamma_{\Omega}) + g_{\mathrm{rot}}(1-\Gamma_{\Omega})+ \Gamma_{\mathrm{E}}(\theta) \, g_{\mathrm{grav}} \, M_{\mathrm{point}}(t) \, f_{\mathrm{OFD}}.
\end{equation}
 
Equation \ref{eq-general} is more general and matches the conditions expected in the upper photosphere and in the wind, and is a function of the local Eddington ratio. Then, an Equation of Motion accounting for the gravity darkening and the oblate distortion of the star, with the variables $u$, $w$ and $w'$, can be derived in a similar way as \cite{cure2004} for a spherical case with an uniform surface brightness, using $g_{\mathrm{tot}}$ given by \ref{eq-general} in the equation of momentum,

\begin{equation}
v \, \frac{dv}{dr}=-\frac{1}{\rho}\frac{dp}{dr} + g_{\mathrm{tot}}(\theta).
\end{equation} 

It is important to note that in our study we use a simpler, but different approximation for $g_{\mathrm{tot}}(\theta)$, i. e., 

\begin{equation}
g_{\mathrm{tot}}(\theta) \approx -g_{\mathrm{grav}}(1-\Gamma_{\Omega}) + g_{\mathrm{rot}}+ \Gamma_{\Omega} \, g_{\mathrm{grav}} \, M_{\mathrm{point}}(t) \, f_{\mathrm{OFD}},
\end{equation} 
 
\noindent where for small $\Gamma_{\Omega}$,  $g_{\mathrm{rot}}$ is not too different from $g_{\mathrm{rot}}(1-\Gamma_{\Omega})$. Our expression is similar to Equation \ref{eq-general} if we assume that $\Gamma_{\mathrm{E}} \approx \Gamma_{\Omega}$ $\ll 1$, as in the cases considered in this study (see Section \ref{grid_hyd}). Finally, please consider that Equation \ref{eq-general} needs to be carefully tested when is applied to cases with considerable values of both $\Gamma_{\mathrm{E}}$ and $\Gamma_{\Omega}$.

\section{Calculation of $\Gamma_{\Omega}$}
\label{gammaO}

\begin{figure}[h]
\center
\includegraphics[width=3 in]{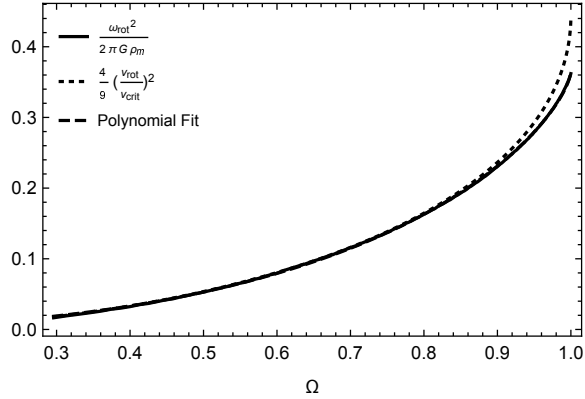}
\caption{Comparison among the ratio ${\omega_{\mathrm{rot}}^{2}}/{2\, \pi \, G\, \rho_{\mathrm{m}}}$ (solid line), the polynomial fit for this ratio (dashed line, overlaid by the solid line) and the term $({4}/{9})\,{v_{\mathrm{rot}}^2(\Omega,\mathrm{eq})}/{v_{\mathrm{crit}}^2}$ (dotted line), all of these terms are dependent on $\Omega$. The polynomial fit and the ratio ${\omega_{\mathrm{rot}}^{2}}/{2\, \pi \, G\, \rho_{\mathrm{m}}}$ have an excellent agreement. In addition, the term $({4}/{9})\,{v_{\mathrm{rot}}^2(\Omega,\mathrm{eq})}/{v_{\mathrm{crit}}^2}$ is a good approximation of ${\omega_{\mathrm{rot}}^{2}}/{2\, \pi \, G\, \rho_{\mathrm{m}}}$ for values of $\Omega$ lower than $\sim 0.9$.}
\label{gamma_factor}
\end{figure}

The definition given for  $\Gamma_{\Omega}$ in the Equations \ref{gammaO1} and \ref{gammaO2} includes the ratio $\omega_{\mathrm{rot}}^{2}/{2\, \pi \, G\, \rho_{\mathrm{m}}}$, which is not easily calculated due to the term $\rho_{\mathrm{m}}$. According to \cite{maeder2000a} this ratio can be expressed in terms of the rotational and critical rotational velocities as

\begin{equation}
\frac{\omega_{\mathrm{rot}}^{2}}{2\, \pi \, G\, \rho_{\mathrm{m}}} = \frac{4}{9}\frac{v_{\mathrm{rot}}^2(\Omega,\mathrm{eq})}{v_{\mathrm{crit}}^2} V'(\Omega) \frac{R_{\mathrm{pole}}^2}{R^2(\Omega, \mathrm{eq})},
\end{equation}

\noindent where $v_{\mathrm{crit}}$ is the classical critical rotational velocity independent of $\Gamma_{\mathrm{E}}$ (Equation \ref{vcrit}). $V'(\Omega)$ is the ratio of the actual volume of a star rotating at $\Omega$ to the volume of a sphere of radius $R_{\mathrm{pole}}$. The term $V'(\Omega) {R_{\mathrm{pole}}^2}/{R^2(\Omega, \mathrm{eq})}$ varies from 1 to 0.813 as  $v_{\mathrm{rot}}$ goes from zero to $v_{\mathrm{crit}}$. For low or moderated velocities this term tends to 1, but  for higher rotational velocities it is necessary to calculate the value  of $V'(\Omega) {R_{\mathrm{pole}}^2}/{R^2(\Omega, \mathrm{eq})}$. To this purpose,  we perform a polynomial fit to the ratio  ${\omega_{\mathrm{rot}}^{2}}/{2\, \pi \, G\, \rho_{\mathrm{m}}}$, whose form is 
\begin{equation}
\frac{\omega_{\mathrm{rot}}^{2}}{2\, \pi \, G\, \rho_{\mathrm{m}}} = 0.0022 + 0.4052 \,x  + 0.1478\, x^{2} - 0.1877\, x^{3}
\label{fit}
\end{equation}
\noindent with $x=\left(\Omega \, {R(\Omega, \mathrm{eq})}/{R_{\mathrm{eq}}^{\mathrm{max}}} \right)^2$. Now, the values of $\Gamma_{\Omega}$ are easily calculated based on this polynomial fit. Figure \ref{gamma_factor} shows the comparison among the ratio ${\omega_{\mathrm{rot}}^{2}}/{2\, \pi \, G\, \rho_{\mathrm{m}}}$, the polynomial fit given in Equation \ref{fit} and the term $ ({4}/{9})\,{v_{\mathrm{rot}}^2(\Omega,\mathrm{eq})}/{v_{\mathrm{crit}}^2}$ as function of $\Omega$. From this Figure, we observe that for values of $\Omega$ higher than $\sim0.9$ the term $ ({4}/{9})\,{v_{\mathrm{rot}}^2(\Omega,\mathrm{eq})}/{v_{\mathrm{crit}}^2}$ ceases to be a good approximation and the polynomial fit has a excellent agreement with the ratio ${\omega_{\mathrm{rot}}^{2}}/{2\, \pi \, G\, \rho_{\mathrm{m}}}$ for the whole range of $\Omega$.

\bibliographystyle{aasjournal}
\bibliography{citas}


\end{document}